\documentclass[apj]{emulateapj}
 \usepackage{apjfonts}

\usepackage{amsmath}

\def\sun{\hbox{$\odot$}}
\newcommand{\msun}{\!\,{\rm M}_\odot}

\def \spose#1{\hbox  to 0pt{#1\hss}}  
\def \lta{\mathrel{\spose{\lower 3pt\hbox{$\sim$}}\raise  2.0pt\hbox{$<$}}}
\def \gta{\mathrel{\spose{\lower  3pt\hbox{$\sim$}}\raise 2.0pt\hbox{$>$}}}

\shorttitle{Stellar-to-Halo Mass Relationship}
\shortauthors{Moster et al.}

\begin{document}

\title{Constraints on the relationship between stellar mass and halo mass at low and high redshift}

\author{Benjamin P. Moster\altaffilmark{1}, Rachel S. Somerville\altaffilmark{1,2}, Christian Maulbetsch\altaffilmark{1}, Frank C. van den Bosch\altaffilmark{1}, Andrea V. Macci\`o\altaffilmark{1},  Thorsten Naab\altaffilmark{3},  and Ludwig Oser\altaffilmark{3}}

\altaffiltext{1}{Max-Planck-Institut f\"ur Astronomie, K\"onigstuhl 17, 69117 Heidelberg, Germany;
   moster@mpia.de, maulbets@mpia.de, vdbosch@mpia.de, maccio@mpia.de.}

\altaffiltext{2}{Space Telescope Science Institute, 3700 San Martin Drive, Baltimore MD 21218;
   somerville@stsci.edu.}

\altaffiltext{3}{Universit\"ats-Sternwarte M\"unchen, Scheinerstr. 1, 81679 M\"unchen, Germany;
   naab@usm.lmu.de, oser@usm.lmu.de.}

\begin{abstract}
We use a statistical approach to determine the relationship between
the stellar masses of galaxies and the masses of the dark matter halos
in which they reside. We obtain a parameterized stellar-to-halo mass (SHM)
relation by populating halos and subhalos in an $N$-body simulation
with galaxies and requiring that the observed stellar mass function
be reproduced. We find good agreement with constraints from galaxy-galaxy
lensing and predictions of semi-analytic models. Using this mapping,
and the positions of the halos and subhalos obtained from the
simulation, we find that our model predictions for the galaxy
two-point correlation function (CF) as a function of stellar mass are in
excellent agreement with the observed clustering properties in the
SDSS at $z=0$. We show that the clustering data 
do not provide additional strong constraints on the SHM function and
conclude that our model can therefore predict clustering as a function
of stellar mass. We compute the conditional mass function, which
yields the average number of galaxies with stellar masses in the range
$m\pm{\rm d}m/2$ that reside in a halo of mass $M$.  We study the
redshift dependence of the SHM relation and show
that, for low mass halos, the SHM ratio is lower
at higher redshift. The derived SHM relation is used
to predict the stellar mass dependent galaxy CF and
bias at high redshift. Our model predicts that not only are massive
galaxies more biased than low mass ones at all redshifts, but the bias
increases more rapidly with increasing redshift for massive galaxies
than for low mass ones. We present convenient fitting functions for
the SHM relation as a function of redshift, the
conditional mass function, and the bias as a function of stellar mass
and redshift.
\end{abstract}

\keywords{cosmology: theory --- dark matter --- galaxies: clusters:
  general --- galaxies: evolution --- galaxies: halos --- galaxies:
  high-redshift --- galaxies: statistics --- galaxies: stellar content
  --- large-scale structure of universe}

\section{Introduction} \label{s:introduction}

In the standard Cold Dark Matter (CDM) paradigm, the formation of
galaxies is driven by the growth of the large-scale structure of the
Universe and the formation of dark matter halos. Galaxies form by the
cooling and condensation of gas in the centers of the potential wells
of extended virialized dark matter halos
\citep{whiterees78, fallefstathiou1980, blumenthal1984}. In this picture,
galaxy properties, such as luminosity or stellar mass,
are expected to be tightly coupled to the depth of the halo potential
and thus to the halo mass.

There are various different approaches to link the properties of
galaxies to those of their halos. A first method attempts to derive
the halo properties from the properties of its galaxy population using
e.g. galaxy kinematics \citep{erickson1987, zaritsky1993, carlberg1996,
more2009a,more2009b}, gravitational lensing \citep{mandelbaum05,
mandelbaum06,cacciato2008}, or X-ray studies \citep{lin2003,linmohr2004}. 

A second approach is to attempt to model the physics that shapes
galaxy formation \emph{ab initio} using either large numerical
simulations including both gas and dark matter \citep{katz1996,
  springel2003} or semi-analytic models (SAMs) of galaxy formation
\citep[e.g.][]{kauffmann1993, cole1994,somerville1999}. In ``hybrid''
SAMs \citep[e.g.][]{croton06,bower2006}, dark matter ``merger trees''
are extracted from a dark matter only N-body simulation, and gas
processes are treated with semi-analytic recipes. An advantage of this
method is that high-resolution N-body simulations can track the
evolution of individual subhalos \citep{klypin1999, springel01} and
thus provide the precise positions and velocities of galaxies within a
halo. However, many of the physical processes involved in galaxy
formation (such as star formation and various kinds of feedback) are
still not well understood, and in many cases simulations are not able
to reproduce observed quantities with high accuracy.

With the accumulation of data from large galaxy surveys over the last
decade, a third method has been developed, which links galaxies to
halos using a statistical approach. The Halo Occupation
Distribution (HOD) formalism specifies the probability distribution
for a halo of mass $M$ to harbour $N$ galaxies with certain intrinsic
properties, such as luminosity, color, or type \citep[e.g.][]{peacock2000,
seljak2000, white2001, berlind2002}. More complex
formulations of this kind of modelling, such as the conditional
luminosity function (CLF) formalism \citep{yang03,vdb2003,yang04} have extended the HOD
approach. These methods have the advantage that they do not rely on
assumptions about the (poorly understood) physical processes that
drive galaxy formation. In this way, it is possible to constrain the
relationship between galaxy and halo properties (and thus, indirectly,
the underlying physics), and to construct mock catalogs that reproduce
in detail a desired observational quantity (such as the luminosity
function). One disadvantage of the classical HOD approach was that one
had to make assumptions about the distribution of positions and
velocities of galaxies within their host halos. In addition, the
results of the HOD modelling can be difficult to interpret in terms of
the underlying physics of galaxy formation.

In recent years, HOD models have been introduced that make use of
information about the positions, velocities and masses of halos and
subhalos extracted from a dissipationless N-body simulation. The
(sub)halo mass is then empirically linked to galaxy properties by
requiring that a statistical observational quantity (e.g. galaxy
luminosity function and/or galaxy two-point-correlation-function) is
reproduced. This is either done by assuming parameterized functions to
relate galaxy properties (such as luminosity) to halo mass or by
assuming a non-parametric monotonic relation. It has been shown that
these simple models reproduce galaxy clustering as a function of
luminosity over a wide range in redshift \citep{kravtsov2004,
tasitsiomi2004,tinker2005,valeostriker06,conroy06, shankar2006,
wang06,marin2008}.

Observationally, it is well known that galaxy clustering is a function
of spatial scale, galaxy properties (such as luminosity and type), and
redshift. Luminous (massive) galaxies are more strongly clustered than
less luminous (less massive) galaxies \citep{norberg2001,norberg2002,
zehavi2002,zehavi2005,li06}. One can split the galaxy two-point correlation
function (2PCF) into two separate parts: the one-halo and the two-halo terms.
The one-halo term, which dominates on small scales, depends strongly on the
galaxy distribution within the halo as well as the details of the HOD.
The two-halo term, which dominates on scales that are much larger than a typical halo, is
proportional to the auto-correlation of the halo population. In general the two terms are not
expected to combine to produce a featureless power-law, but generally show a
break or dip at the scale where the transition from the one-halo to the two-halo
term occurs \citep{zehavi2004}.

The extensive multi-wavelength spectrophotometric information that is
now available for large numbers of galaxies allows us to estimate
physical parameters of galaxies, such as stellar masses, instead of
relying on observational properties such as magnitudes \citep{bell2001,
kauffmann03,panter2004}. These estimates can even be
obtained --- with a proper measure of caution --- for high redshift
galaxies. Stellar mass estimates have been presented in the literature
for galaxies up to redshifts as high as $z\sim 6$
\citep{yan2006,eyles2007}, and stellar
mass function estimates have been presented up to $z\sim 5$
\citep{drory2005,fontana2006,elsner2008}.
The goal of our paper is to develop a ``Conditional Stellar
Mass Function'' (CMF) formalism, which is the stellar mass analog of the
CLF. The CMF yields the average number of galaxies with stellar masses
in the range $m\pm{\rm d}m$ as a function of the host halo mass $M$
and can be regarded as the stellar mass function (SMF) for halos of mass
$M$. We apply this formalism at low redshift and up to the highest
redshifts where reliable observational stellar mass estimates are
available ($0.1 \lesssim z \lesssim 4$). In this way, we derive a parameterized
relationship between dark matter halo mass and galaxy mass as a
function of redshift.

Using a parameterized relationship has several advantages. First, it
provides a convenient way for other researchers to make use of our
results and obtain an expression for stellar mass as a function of
halo mass. Second, it is straightforward to include scatter in the
relation, which is physically more realistic: one just has to choose a
number drawn from an assumed random distribution and add that to the
average relation. Finally, it is straightforward to treat central and
satellite galaxies separately and assume different relations between
stellar and halo mass for those populations.  However, here we make
the assumption that both populations follow the same relation, which
has consequences for the clustering predictions of our model.

Using the CMF derived \emph{only} from constraints from the observed
SMF, we compute the predicted (projected) galaxy CF at $z\sim 0$ as a
function of stellar mass, and find good agreement with the
observational results of \citet{li06}.
Furthermore, we show that assuming central and satellite galaxies
follow the same relation between stellar and halo mass, adding the
clustering constraints does not tighten the constraints on
our model parameters; i.e., any model that satisfies the mass function
constraints will produce the correct clustering. Based on this result,
we use our redshift-dependent CMF results to \emph{predict} the
clustering as a function of stellar mass and redshift.
To date, observational measurements of clustering as a function of stellar mass
have only been published for $z\lta 1$ \citep{meneux2008,meneux2009}.
We show that our model predictions agree very well with these measurements.
Very soon it will be possible to test our predictions for redshifts beyond
$z=1$ with the results from deep wide-field surveys (e.g. MUSYC, UKIDDS, etc).
We again present convenient fitting functions for the galaxy bias as a function
of both stellar mass and redshift. In a companion paper we will employ our
estimates of galaxy bias in order to compute the ``cosmic variance'',
the uncertainty in observational estimates of the volume density of
galaxies arising from the underlying large-scale density fluctuations.

This paper is organized as follows: in section \ref{s:simulation} we
describe the $N$-body simulation, the halo finding algorithm that was
used to obtain a halo catalogue and the treatment of `orphaned'
galaxies. Section \ref{s:galaxies} specifies our model: we motivate
the form of the stellar-to-halo mass (SHM) relation and constrain it by
requiring that the observed SMF is reproduced. The
clustering properties of galaxies are then inferred from those of the
halo population. We discuss the meaning of the parameters of the
SHM relation and demonstrate that clustering puts
only weak constraints on them. In section \ref{s:cmf} we introduce the
CMF, which describes how halos are occupied by galaxies, and compute
the occupation numbers. Section \ref{s:comparison} gives a comparison
between our results and several other models and observations. In
section \ref{s:redshift} we apply our method to higher redshifts and
determine the redshift dependence of the SHM
relation. We make predictions of the stellar mass dependent galaxy
CF at higher redshift which we use to compute the
galaxy bias. Finally, we summarize our methods and conclusions in
section \ref{s:conclusions}.

Throughout this paper we assume a $\Lambda$CDM cosmology with
($\Omega_m$,$\Omega_{\Lambda}$,$h$,$\sigma_8$,$n$) =
($0.26$,$0.74$,$0.72$,$0.77$,$0.95$). We employ a \citet{kroupa01}
initial mass function (IMF) and convert all stellar masses to this
IMF. In order to simplify the notation we will use the capital $M$ to
denote dark matter halo masses and the lower case $m$ to denote galaxy
stellar masses.

\section{The simulation and halo catalogs} 
\label{s:simulation}

High-resolution dissipationless N-body simulations have shown that
distinct halos\footnote{We refer to virialized halos that are not
subhalos of another halo as ``distinct''.} contain subhalos which
orbit within the potential of their host halo. These subhalos were
distinct halos in the past, and entered the larger halo via merging
during the process of hierarchical assembly. We will refer to the
galaxy at the center of a distinct halo as a central galaxy, and the
galaxies within subhalos as ``satellites'', and we will use the term
`halo' to refer to the distinct halo for central galaxies and to the
subhalo in which the galaxy originally formed for satellite galaxies.

{\emph Ab initio} models of galaxy formation predict that the stellar
mass of a galaxy is tightly correlated with the depth of the potential
well of the halo in which it formed. For distinct halos, the relevant
mass is the virial mass at the time of observation. Subhalos, however,
lose mass while orbiting in a larger system as their outer regions are
tidally stripped. Stars are centrally concentrated and more tightly
bound than the dark matter, however, and so the stellar mass of a
galaxy which is accreted by a larger system probably changes only
slightly until most of the dark matter has been stripped
off. Therefore the subhalo mass at the time of observation is probably
not a good tracer for the potential well that shaped the galaxy
properties. A better tracer is the subhalo mass at the time that it
was accreted by the host halo, or its maxmimum mass over its
history\footnote{In an idealized situation, halo mass should increase
  monotonically with time until the halo becomes a subhalo, at which
  point the mass begins to decrease due to tidal stripping.}.
This was first proposed by \citet{conroy06}.

The population of dark matter halos used in this work is drawn from an
$N$-body simulation run with the simulation code {\small GADGET}-2
\citep{springel05a} on a SGI AltixII at the University Observatory Munich.
The cosmological parameters of the simulation are
chosen to match results from {\small WMAP}-3 \citep{spergel06} for a
flat $\Lambda$CDM cosmological model: $\Omega_m=0.26$,
$\Omega_{\Lambda}=0.74$, $h=H_0/(100$~km~s$^{-1}$~Mpc$^{-1})=0.72$,
$\sigma_8=0.77$ and $n=0.95$.  The initial conditions were generated
using the GRAFIC  software package \citep{bertschinger2001}. The
simulation was done in a periodic box with side length $100$ Mpc,
and contains $512^3$ particles with a particle mass of
$2.8\times 10^8\msun$ and a force softening of $3.5$ kpc.

Dark matter halos are identified in the simulation using a
friends-of-friends (FoF) halo finder.  Substructures inside the FoF
groups are then identified using the {\small SUBFIND} code described
in \cite{springel01}. For the most massive subgroup in a FoF group the
virial radius and mass are determined with a spherical overdensity
criterion: the density inside a sphere centered on the most bound
particle is required to be greater than or equal to the value
predicted by the spherical collapse model for a tophat perturbation in
a $\Lambda$CDM cosmology \citep{bryan98}.  As discussed above, for
subhalos we use the maximum mass over its past history, which is
typically the mass when the halo was last a distinct halo and did not yet
overlap with its later host.  Merger trees were constructed out of the
halo catalogs at 94 time-steps, equally spaced in expansion factor
($\Delta a=0.01$), based on the particle overlap of halos at different
time-steps. 

Due to the finite mass resolution of the simulation
($M_{\rm min,halo}\simeq 10^{10} \msun$), subhalos can no longer be
identified when their mass has dropped below this limit due to tidal
stripping.  Since mass loss can be substantial (>90\%) this is
important even for fairly massive subhalos. A special treatment of
these so-called ``orphans'' is necessary. We determine the orbital
parameters at the last moment when a subhalo is identified in the
simulation and use them in the dynamical friction recipe of
\cite{boylan08}, which is applicable at radii $r<r_{vir}$. We
also tried an alternate recipe in which we make no explicit use of the
subhalo information, but apply the dynamical friction formula from the
time when the satellites first enter the host halo. We obtained very
similar subhalo mass functions and radial distributions with the
alternate recipe, confirming the self-consistency of the approach.

For the halo positions in the determination of CFs,
we use the coordinates of the most bound particle for distinct and
subhalos. For orphans, by definition, the position is not known, so we
follow the position of the most bound particle from the last time-step
when a subhalo was identified. Since the dynamical friction force
vanishes in the dark matter only simulation after a subhalo is dissolved,
yet not in reality when a galaxy is present at the center of the subhalo, the
distance to the center of the host halo might be slightly
overestimated with this prescription.

\section{Connecting galaxies and halos} \label{s:galaxies}

In this section we describe how we derive the relationship connecting
the stellar mass of a galaxy to the mass of its dark matter halo.  In
the standard picture of galaxy formation, gas can only cool and form
stars if it is in a virialized gravitationally bound dark matter halo
\citep{whiterees78}. In this model the gas cooling rate, the star
formation rate and thus the properties of the galaxy depend
mainly on the virial mass of the host halo. Thus we expect the stellar
mass of a central galaxy to be strongly correlated with the virial mass of
the halo in which the galaxy formed. As we discussed in the last section,
this corresponds to the virial mass for central galaxies, and to the
maximum mass over the halo's history for satellite galaxies. In the
rest of this work, unless noted otherwise, the halo mass $M$ will
represent:
\begin{equation}
M = \begin{cases}
          M_{\rm vir} & \text{for host halos}\\
          M_{\rm max} & \text{for subhalos}
       \end{cases}
\label{eqnmmax}
\end{equation} 

Note that we have also experimented with instead using the present
mass for subhalos, and found that we were not able to reproduce the
galaxy clustering properties \citep[see also][]{conroy06}.

\subsection{The stellar-to-halo mass relation}\label{s:galaxymap}

\begin{figure}
\centering
\plotone{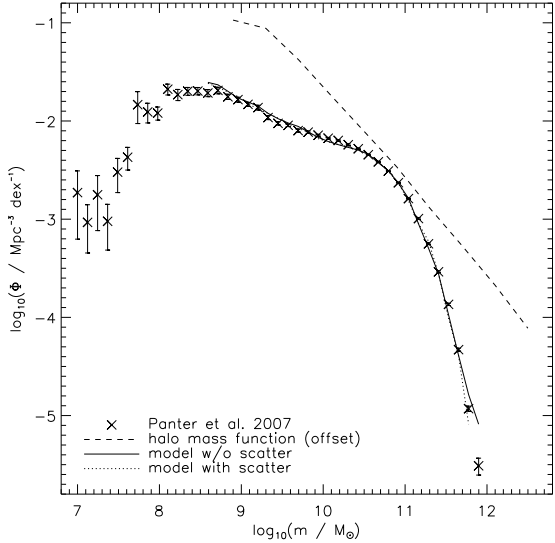}
\caption[Comparison between the halo mass function, observed and model
  galaxy mass functions]{A comparison between the halo mass function
  offset by a factor of 0.05 (dashed line), the observed galaxy mass
  function (symbols), our model without scatter (solid line) and our model
  including scatter (dotted line).We see that the halo
  and the galaxy mass functions are different shapes,
  implying that the stellar-to-halo mass ratio $m/M$ is not
  constant. Our four parameter model for the halo mass dependent
  stellar-to-halo mass ratio is in very good agreement with the
  observations (both including and neglecting scatter).}
\label{f:fig1}
\end{figure}

In order to link the stellar mass of a galaxy $m$ to the mass of its
dark matter halo $M$ we need to specify the SHM
ratio. A direct comparison of the halo mass function $n(M)$ and the
galaxy mass function $\phi(m)$ helps to constrain the stellar-to-halo
mass function. If we assume that every host (sub) halo contains
exactly one central (satellite) galaxy and that each system has
exactly the same SHM ratio $m/M$, the galaxy stellar
mass function can be derived trivially from the halo mass function and
has the same features. The galaxy mass function derived for $m/M =
0.05$ is compared to the observed SDSS galaxy mass function in Figure
\ref{f:fig1}. The observed galaxy mass function is steeper for high
masses and shallower for low masses than the one derived from the halo
mass function. Thus, for a constant SHM ratio there
will inevitably be too many galaxies at the low and high mass end.

This implies that the actual SHM ratio $m/M$ is not
constant, but increases with increasing mass, reaches a maximum around
$m^*$ and then decreases again. Hence we adopt the following
parametrization, similar to the one used in \citet{yang03}:
\begin{equation}
 \frac{m(M)}{M} = 2 \left( \frac{m}{M}\right)_0 \left[\left(\frac{M}{M_1}\right)^{-\beta} + \left(\frac{M}{M_1}\right)^{\gamma}\right]^{-1}
\label{eqnmmap}
\end{equation} 
It has four free parameters: the normalization of the stellar-to-halo
mass ratio $(m/M)_0$, a characteristic mass $M_1$, where the
SHM ratio is equal to $(m/M)_0$, and two slopes
$\beta$ and $\gamma$ which indicate the behavior of $m/M$ at the low
and high mass ends respectively. We use the same parameters for the
central and satellite populations, since -- unlike luminosity -- the
stellar mass of satellites changes only slightly after they are
accreted by the host halo.

Note that though both $\beta$ and $\gamma$ are expected to be
positive, they are not restricted to be so. The SHM
relation is therefore not necessarily monotonic.

\subsection{Constraining the free parameters}
\label{constraining}
Having set up the model we now need to constrain the four free
parameters $M_1$, $(m/M)_0$, $\beta$ and $\gamma$. To do this, we
populate the halos in the simulation with galaxies. The stellar masses
of the galaxies depend on the mass of the halo and are derived
according to our prescription (equation \ref{eqnmmap}). The positions
of the galaxies are given by the halo positions in the $N$-body
simulation.

Once the simulation box is filled with galaxies, it is straightforward
to compute the SMF $\Phi_{mod}(m)$. As we want to
fit this model mass function to the observed mass function
$\Phi_{obs}(m)$ by \citet{panter2007}, we choose the same stellar mass
range ($10^{8.5}-10^{11.85}~\msun$) and the same binsize. The observed
SMF was derived using spectra from the Sloan Digital
Sky Survey Data Release 3 (SDSS DR3); see \citet{panter2004} for a description
of the method.

Furthermore it is possible to determine the stellar mass dependent
clustering of galaxies. For this we compute projected galaxy
CFs $w_{p,mod}(r_p,m_i)$ in several stellar mass
bins which we choose to be the same as in the observed projected
galaxy CFs of \citet{li06}. These were derived using
a sample of galaxies from the SDSS DR2 with stellar masses estimated
from spectra by \citet{kauffmann03}.

We first calculate the real space CF $\xi(r)$. In a
simulation this can be done by simply counting pairs in distance bins:
\begin{equation}
\label{corrfunc}
\xi(r_i) = \frac{dd(r_i)}{N_p(r_i)}-1
\end{equation}
where $dd(r_i)$ is the number of pairs counted in a distance bin and
$N_p(r_i) = 2 \pi N^2 r_i^2 \Delta r_i / L_{\rm box}^3$ where $N$ is the
total number of galaxies in the box. The projected CF
$w_p(r_p)$ can be derived by integrating the real space correlation
function $\xi(r)$ along the line of sight:
\begin{equation}
w_p(r_p) = 2 \int_{0}^{\infty} {\rm d}r_{||} \xi(\sqrt{r_{||}^2+r_p^2}) = 2 \int_{r_p}^{\infty} {\rm d}r \frac{r~\xi(r)}{\sqrt{r^2-r_p^2}} \;,
\end{equation}
where the comoving distance ($r$) has been decomposed into components
parallel ($r_{||}$) and perpendicular ($r_p$) to the line of
sight. The integration is truncated at $45$~Mpc. Due to the finite
size of the simulation box ($L_{\rm box}=100$~Mpc) the model correlation
function is not reliable beyond scales of
$r\sim0.1~L_{\rm box}\sim10$~Mpc.

In order to fit the model to the observations we use Powell's
directions set method in multidimensions (e.g. Press et al. 1992) to
find the values of $M_1$, $(m/M)_0$, $\beta$ and $\gamma$ that
minimize either $$ \chi_r^2 = \chi_r^2 (\Phi)= \frac{\chi^2(\Phi)}{N_{\Phi}}$$ (mass
function fit) or $$ \chi_r^2 = \chi_r^2 (\Phi)+\chi_r^2 (w_p) = \frac{\chi^2(\Phi)}{N_{\Phi}} +
\frac{\chi^2(w_p)}{N_r\;N_m}$$ (mass function and
projected CF fit)
with $N_{\Phi}$ and $N_{r}$ the number of data points
for the SMF and projected CFs, respectively, and $N_{m}$ the number
of mass bins for the projected CFs.

In this context $\chi^2(\Phi)$
and $\chi^2(w_p)$ are defined as:
\begin{eqnarray}
\chi^2(\Phi) &=& \sum_{i=1}^{N_{\Phi}} \left[ \frac{\Phi_{\rm mod}(m_i)-\Phi_{\rm obs}(m_i)}{\sigma_{\Phi_{\rm obs}(m_i)}} \right]^2\notag\\
\chi^2(w_p) &=& \sum_{i=1}^{N_m} \sum_{j=1}^{N_r} \left[ \frac{w_{p,\rm mod}(r_{p,j},m_i)-w_{p,\rm obs}(r_{p,j},m_i)}{\sigma_{w_{p,\rm obs}(r_{p,j},m_i)}} \right]^2\notag\;,
\end{eqnarray}
with $\sigma_{\Phi_{\rm obs}}$ and $\sigma_{w_{p,\rm obs}}$ the errors for the
SMF and projected CFs, respectively.
Note that for the simultaneous fit, by adding the reduced $\chi_r^2$,
we give the same weight to both data sets.

\begin{figure*}
\centering
\epsscale{1.15}
\plotone{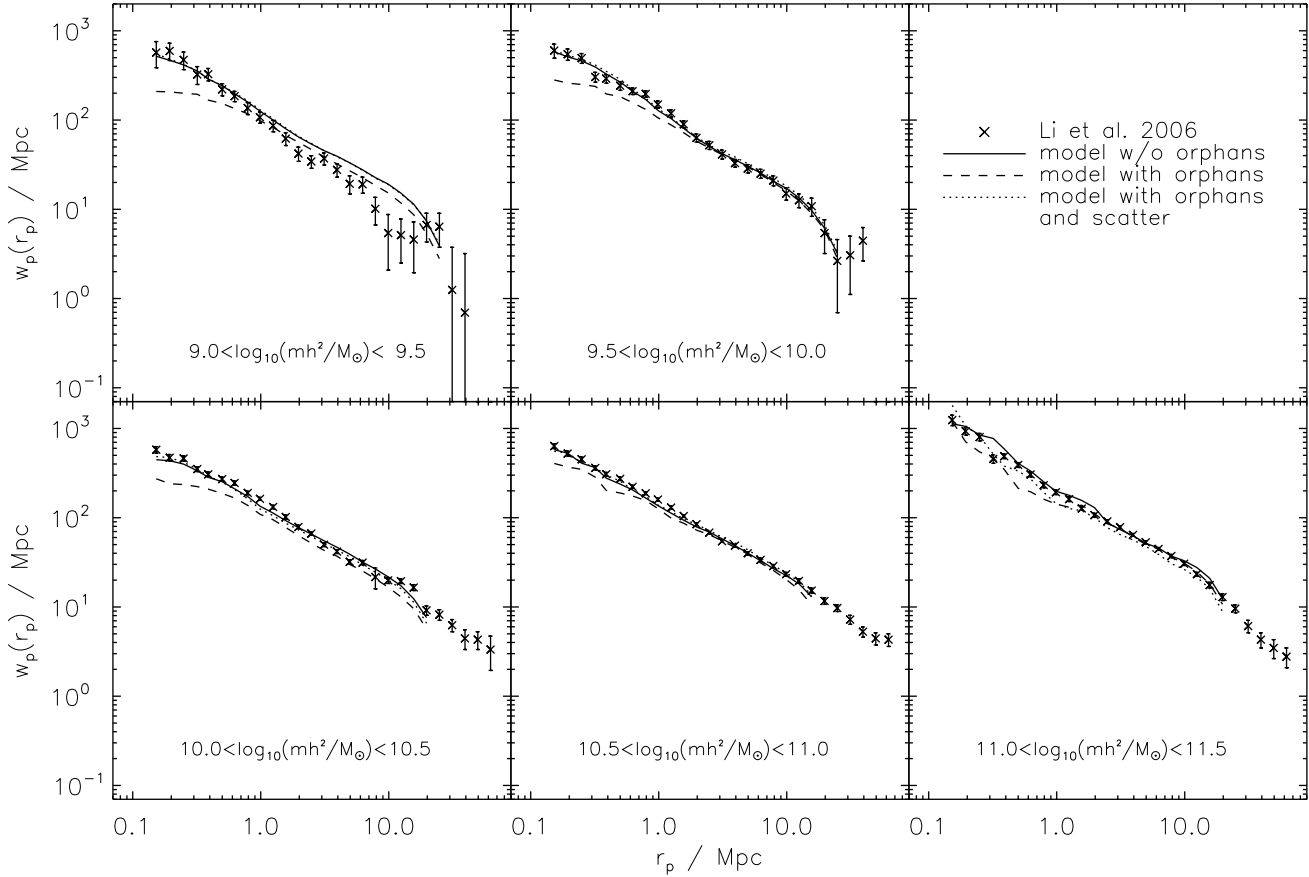}
\caption{Comparison between the model (lines) and observed (symbols
  with errorbars) projected correlation functions. We show the model
  results both including (solid) and excluding (dashed) orphan
  galaxies. The models have been derived by fitting to the stellar mass
  function only.}
\label{f:fig2}
\end{figure*}

\subsection{Estimation of parameter errors}\label{s:probdis}

In order to obtain estimates of the errors on the parameters, we need
their probability distribution prob($A\vert I$), where $A$ is the
parameter under consideration and $I$ is the given background
information. The most likely value of $A$ is then given by: $A_{\rm best}
={\rm max(prob(}A\vert I))$.

As we have to assume that all our parameters are coupled, we can only
compute the probability for a given set of parameters. This
probability is given by:
$${\rm prob(}M_1,(m/M)_0,\beta,\gamma\vert I) \propto \exp(-\chi^2)$$
In a system with four free parameters $A,B,C$ and $D$ one can
calculate the probability distribution of one parameter (e.g. $A$) if
the probability distribution for the set of parameters is known, using
marginalization:
\begin{eqnarray}
{\rm prob(}A\vert I) &=& \int_{-\infty}^{\infty}{\rm prob(}A,B\vert I) {\rm d}B\notag\\
&=& \int_{-\infty}^{\infty}{\rm prob(}A,B,C,D\vert I) {\rm d}B {\rm d}C {\rm d}D\notag
\end{eqnarray}

Once the probability distribution for a parameter is determined, one
can assign errors based on the confidence intervals. This is the
shortest interval that encloses a certain percentage $X$ of the area
under the posterior probability distribution. For the 1-sigma error $X = 68\%$
while for the 2-sigma error $X = 95\%$. Assuming that the probability distribution
has been normalized to have unit area we seek $A_1$ and $A_2$ such that

$$ \int_{0}^{A_1} {\rm prob(}A\vert I) {\rm d}A = \int_{A_2}^{\infty}
{\rm prob(}A\vert I) {\rm d}A = \frac{1-X}{2} .$$

Finally the parameter $A$ is given as $A = A_{\rm best}\text{
}_{-\sigma_{-}}^{+\sigma_{+}}$ with $\sigma_{+} = A_2 - A_{\rm best} $
and $\sigma_{-} = A_{\rm best} - A_1$.
The errors derived in this way only include sources that have been considered
when computing $\chi^2$. The calculation of the errors applies for uncorrelated
data points. Since in our case the data points are correlated the values of the errors
are slightly modified. Also errors caused by cosmic variance are not included.

\section{Fitting results}\label{s:results}

Here we present the results we obtain by fitting to the stellar mass
function only, and for the combined fit to the SMF
and the projected CF.

\subsection{The stellar mass function fit}

\begin{deluxetable}{lllllll}[hb!]
\tablecaption{Fitting results for Stellar-to-Halo Mass relationship}
\tablehead{
  \colhead{} &
  \colhead{$\log M_1$} &
  \colhead{$(m/M)_0$} &
  \colhead{$\beta$} &
  \colhead{$\gamma$} &
  \colhead{$\chi_r^2(\Phi)$} &
  \colhead{$\chi_r^2(w_p)$}
}
\startdata
best fit & 11.884  & 0.02820 & 1.057 & 0.556 & 1.56 & 3.83\\
$\sigma^+$ &  ~~0.030 & 0.00061 & 0.054 & 0.010 & &\\
$\sigma^-$ &  ~~0.023 & 0.00053 & 0.046 & 0.004 & &\\
\enddata
\tablecomments{No scatter included. All masses are in units of $\msun$}
\label{t:mfresults}
\end{deluxetable} 

First we fit to the SDSS SMF and use the derived
best-fit parameters to calculate the model projected correlation
functions. Note that for now, we do not take into account
any possible scatter in the $m(M)$ relation. We will consider scatter
in \S\ref{s:scattersec}.

We see in Figure \ref{f:fig1} that our fit produces excellent
agreement with the observed SMF. Using the approach
described above we also compute the errors on the parameters. The
results are summarized in Table \ref{t:mfresults}.

Having derived the best-fit parameters, we can predict the projected
CFs. We present the results both including and not
including orphan galaxies, where we have fitted to the SMF for each case.

Figure \ref{f:fig2} shows a comparison between our model and the SDSS
projected correlation functions in five stellar mass bins ranging from
$\log m/\msun=9.0$ to $\log m/\msun=11.5$ with a binsize of $0.5 {\rm~dex}$.
The correlation function that has been derived without orphans is too low at
small scales and can be regarded as a lower limit. Neglecting these galaxies
results in an underprediction of satellite galaxy clustering. As on
small scales the projected CF depends mainly on
the one-halo term this results in the underprediction of $w_p(r_p)$.
This effect weakens for the clustering of more massive galaxies as they are
more likely to be central galaxies and thus not effected by tidal stripping at all.

The agreement with the observationally derived $w_p(r_p)$ for the
catalogue including orphaned galaxies is very good, which is also
reflected in the low value of $\chi_r^2(w_p) = 3.83$. Note that this
value has been calculated with the parameters from the mass function
fit given above and does not correspond to a fit to the projected CFs.

Note that we plot the projected CFs only up to $20\, {\rm
Mpc}$. Because of the finite box size, the clustering of host
halos and thus central galaxies is underpredicted at large scales
independent of mass.  Additionally, due to the lack of
long-wavelength modes, massive halos and galaxies can be
underproduced leading to an underprediction of $w_p$ for the massive
objects, independent of scale. However, the latter effect is very
small, since the abundance of the massive halos in our simulation
agrees very well with the predicted average \citep{sheth1999}.

As a test we also used the present mass instead of the maximum mass for
subhalos.  We then found that the projected CF was
underpredicted particularly on small scales. This effect is due to
tidal stripping of subhalos and is thus strongest at small scales
where the subhalo contribution dominates.

\subsection{The combined fit}\label{combinedfit}

\begin{figure}
\centering
\plotone{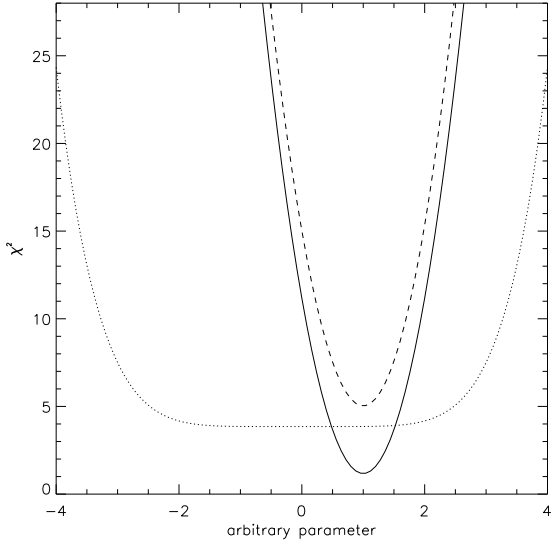}
\caption{Sketch of the probability distributions for a simultaneous
  fit. The solid line corresponds to $\chi^2(m)$ and the dotted line
  to $\chi^2(w_p)$. The dashed line is the sum of both. Since
  $\chi^2(w_p)$ is flat at the minimum, $\chi^2_{\rm tot}$ follows
  $\chi^2(m)$ with an offset. The resulting probability distribution
  does not change (after normalization).}
\label{f:fig3}
\end{figure}

We now investigate whether we can improve the agreement between the
model and the observed projected CFs by performing a
combined fit as described above. We obtain the same parameters as
those we derived from the fit to the SMF alone. This
seems surprising, but on further inspection we find that this is due
to $\chi^2(m)$ being a lot more sensitive to changes of the parameters
than $\chi^2(w_p)$. This means that if one changes the parameters a
little in order to improve the fit to the projected correlation
functions, one can get a slightly better agreement between the model
and the observed projected CFs only at the cost of a
large disagreement between the model and the observed stellar mass
functions. In other words: $\chi^2(w_p)$ is much flatter around its
minimum than $\chi^2(w_p)$, as shown in Figure \ref{f:fig3}.

This means that, assuming that both central and satellite
galaxies follow the same SHM relation, the model that matches the
SMF can reproduce the correct clustering. However, if subhalos have
a different SHM ratio there is an infinite number of solutions that
match the SMF but produce very different correlation functions. The
only way to constrain the SHM relations then is to take the
clustering data into account. By adopting different SHM relations
for central and satellite populations it is even possible to produce
a slightly better fit to the correlation functions \citep{wang06}.

On the other hand, if one wants to {\em predict} clustering as a
function of stellar mass (e.g. at higher redshift) then one has to
make an assumption about how the SHM ratios of central and satellite
galaxies are related. We made the very simple assumption, that the
relation between the stellar mass of central galaxies and the virial
mass of their host halo and the relation between the stellar mass of
satellite galaxies and the mass of the subhalo at the time of
accretion is the same, and have shown that this leads to very good
predictions for the mass dependent clustering.  We conclude that under
this simple assumption we can use our model to predict clustering as a
function of stellar mass.

\subsection{The resulting stellar-to-halo mass relation} \label{masstomass}

\begin{figure}
\centering
\epsscale{1.15}
\plotone{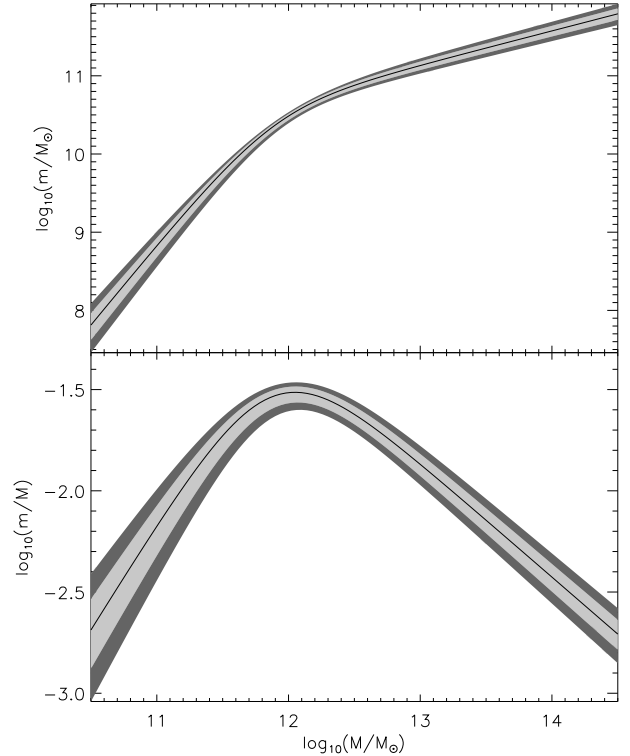}
\caption[The derived relation between stellar mass and halo mass]{The
  derived relation between stellar mass and halo mass. The light
  shaded area shows the $1\sigma$-region while the dark and light
  shaded areas together show the $2\sigma$-region.
  The upper panel shows the SHM relation while the lower panel shows the SHM ratio.}
\vspace{0.5cm}
\label{f:fig4}
\end{figure}

The upper panel of Figure \ref{f:fig4} shows the derived stellar mass
as a function of halo mass.
The light shaded area gives the 68\% confidence
interval while the dark and light shaded areas together give the
95\% confidence interval. These have been derived using a set
of different models computed on a mesh, as described in
\S\ref{s:probdis}.

For the SHM ratio we apply the same
procedure. The result is shown in the lower panel of Figure
\ref{f:fig4}. We see that the SHM ratio has the form
we expected: it increases with increasing halo mass, reaches its
maximum value around $M_1$ and then decreases again.

\subsection{Meaning of parameters and correlations}\label{s:parameters}

\begin{figure}
\centering
\epsscale{1.2}
\plotone{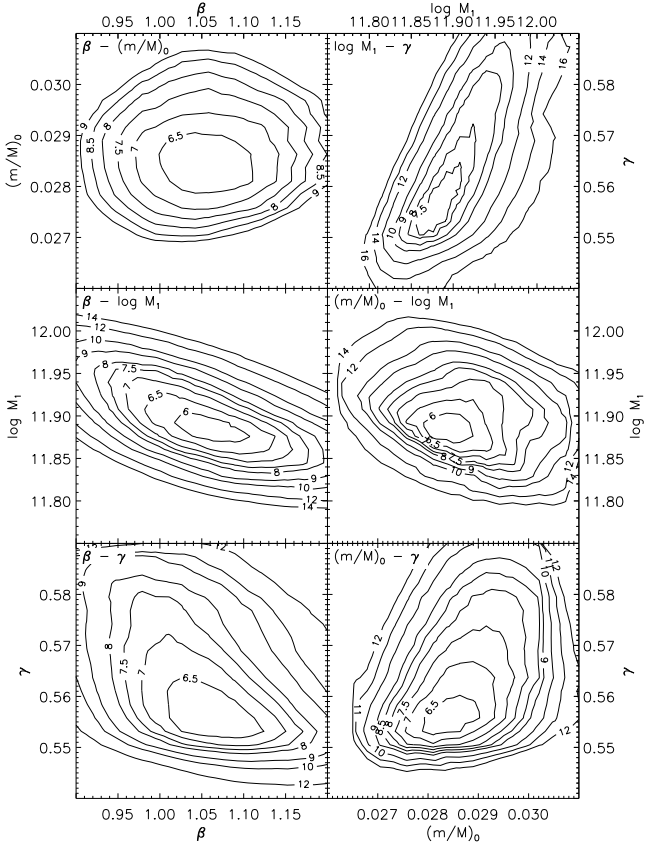}
\caption{Correlations between the model parameters. The panels show
  contours of constant $\chi^2$ (i.e. constant probability) for
    the fit including contraints from the SMF only. The parameter
  pairs are indicated in each panel.}
\vspace{0.5cm}
\label{f:fig5}
\end{figure}

We now explore the effects of changing each parameter in order to
understand how they affect the SMF.
\label{ratio}
If we keep $M_1$, $\beta$ and $\gamma$ fixed and only vary $(m/M)_0$,
this corresponds to changing the stellar mass of the galaxy that lives
inside each halo by a constant factor. This has no impact on the form
of the SMF. Its shape stays the same, while only the position on the
stellar mass axis changes. Due to the monotonic form of the SMF this
directly determines the value of the normalization $\phi^*$. For a
larger value of $(m/M)_0$ we get a larger value of $\phi^*$.

\label{m1}
Varying only $M_1$ we find that the shape of the SMF
changes drastically. For a higher $M_1$ than our best fit value, we
get too many massive galaxies and too few low mass galaxies, while for
a lower value of $M_1$ we get too few massive galaxies and too many
low mass galaxies. This is because $M_1$ is the characteristic mass
corresponding to the highest SHM ratio. In the
SMF, this corresponds to the knee and we get a
SMF which has its knee at the stellar mass
corresponding to $M_1$. For a larger $M_1$ the knee is shifted to a
higher stellar mass. Together, $M_1$ and the maximum stellar-to-halo
mass ratio $(m/M)_0$ determine the normalization of the stellar mass
function $\phi$ and the characteristic mass $m^*$.

\label{beta}
Changing $\beta$ affects mainly the low mass slope of the stellar mass
function. For larger values of $\beta$ the slope becomes shallower.
As $\beta$ influences mainly the slope of the low mass
end of the SMF, it is strongly related to the
parameter $\alpha$ of the Schechter function. A small value of $\beta$
corresponds to a high value of $\alpha$.

\label{gamma}
If we change $\gamma$, this mainly impacts the slope of the massive
end of the SMF. For larger values of $\gamma$ than for its best-fit
value the slope of the massive end becomes steeper. As $\gamma$
affects mainly the slope of the massive end of the SMF it is not
coupled to a parameter of the Schechter function though it is related
to the high-mass cutoff, assumed to be exponential in a Schechter
function.

Figure \ref{f:fig5} shows the contours of the two-dimensional probability
distributions for the parameters pairs. We see a
correlation between the parameters [$M_1,\gamma$] and
[$(m/M)_0,\gamma$] and an anti-correlation between [$\beta,\gamma$],
[$\beta,M_1$] and [$(m/M)_0,M_1$]. There does not seem to be a
correlation between [$\beta,(m/M)_0$].

\subsection{Introducing scatter}\label{s:scattersec}

\begin{figure}
\centering
\epsscale{1.15}
\plotone{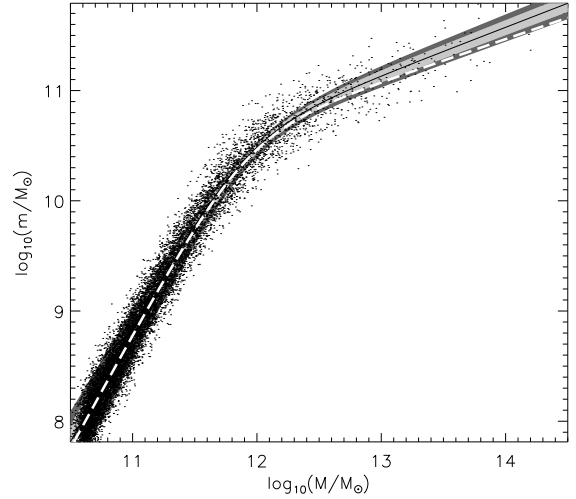}
\caption{Stellar mass as a function of halo mass with
  $\sigma_m=0.15\, {\rm dex}$. The solid line corresponds to our model
  without scatter while the points represent the model with scatter
  (note that only 20\% of the total number of objects are
  plotted). The relation between halo mass and the average stellar
  mass for the model with scatter is shown by the dashed line.}
\label{f:fig6}
\end{figure}

Up until now we have assumed that there is a one-to-one, deterministic
relationship between halo mass and stellar mass. However, in nature,
we expect that two halos of the same mass $M$ may harbor galaxies with
different stellar masses, since they can have different halo concentrations,
spin parameters and merger histories.

For each halo of mass $M$, we now assign a stellar mass $m$ drawn from
a log-normal distribution with a mean value given by our previous
expression for $m(M)$ (Equation \eqref{eqnmmap}), with a variance of
$\sigma_m^2$. We assume that the variance is a constant for all halo
masses, which means that the percent deviation from $m$ is the same
for every galaxy. This is consistent with other halo occupation
models, semi-analytic models and satellite kinematics
\citep{cooray2006, vdb2007,more2009b}.

Assuming a value of $\sigma_m = 0.15{\rm~dex}$ and fitting the stellar
mass function only, we find the values given in Table
\ref{t:mfscatresults}.  These values lie within the (2$\sigma$) error
bars of the best-fit values that we obtained with no scatter. The
largest change is on the value of $\gamma$, which controls the slope
of the SHM relation at large halo masses.
The SMF and the projected CFs for the model including scatter are
shown in Figures \ref{f:fig1} and \ref{f:fig2}, respectively, and show
very good agreement with the observed data.

\begin{deluxetable}{lllllll}[hb!]
\tablecaption{Fitting Results for for Stellar-to-Halo Mass relationship}
\tablehead{
  \colhead{} &
  \colhead{$\log M_1$} &
  \colhead{$(m/M)_0$} &
  \colhead{$\beta$} &
  \colhead{$\gamma$} &
  \colhead{$\chi_r^2(\Phi)$} &
  \colhead{$\chi_r^2(w_p)$}
}
\startdata
best fit          & 11.899  & 0.02817 & 1.068 & 0.611 & 1.42 & 4.21\\
$\sigma^+$ &  ~~0.026 & 0.00063 & 0.051 & 0.012 & &\\
$\sigma^-$  &  ~~0.024 & 0.00057 & 0.044 & 0.010 & &\\
\enddata
\tablecomments{Including scatter $\sigma_m=0.15$. All masses are in units of $\msun$}
\label{t:mfscatresults}
\end{deluxetable}

In Figure \ref{f:fig6} we compare our model without scatter with the
model including scatter. We have also included the relation between
halo mass and the average stellar mass. Especially at the massive end
scatter can influence the slope of the SMF, since
there are few massive galaxies. This has an impact on $\gamma$ and as
all parameters are correlated scatter also affects the other
parameters. We thus see a difference between the model without scatter
and the most likely stellar mass in the model with scatter in Figure
\ref{f:fig6}.

\section{The conditional mass function} \label{s:cmf}
\begin{figure*}
\centering
\epsscale{1.11}
\plotone{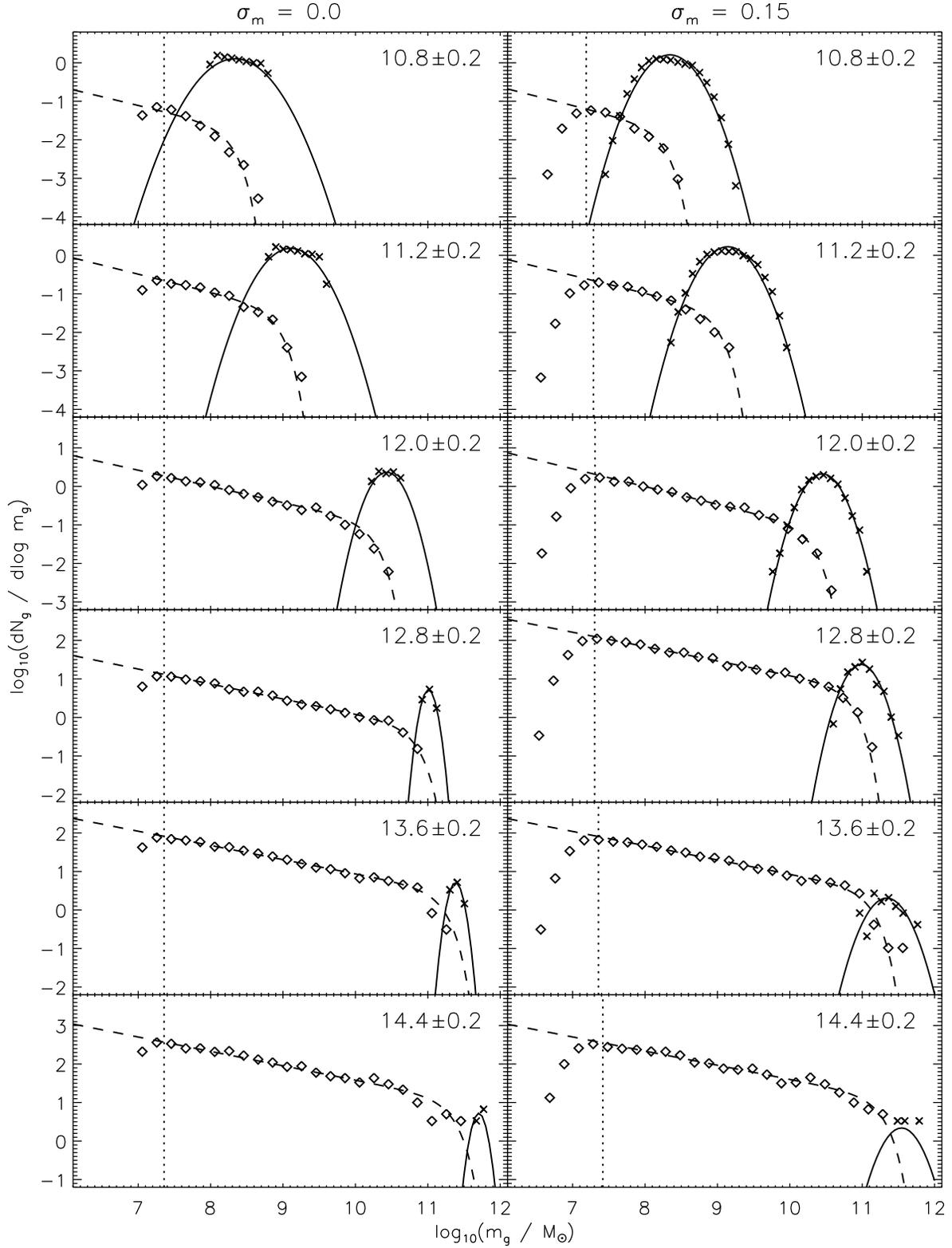}
\caption{The conditional mass function (CMF) predicted by our model at
  $z=0$. We plot the derived SMFs (${\rm d}\tilde{n}_g / {\rm d}\log
  m$) in a subsample of halo mass bins. The left panels show the CMF
  for a model without scatter while the right panels show the CMF with
  scatter of $\sigma_m=0.15$. The label in each panel is the range of
  host halo mass $\log M/\msun$. The stellar mass functions are
  normalized such that a host halo contains exactly one central
  galaxy. The total CMF consists of a central galaxy part (crosses)
  and a satellite part (diamonds). The central part is described by a
  lognormal distribution (solid line) and the satellite part is
  described by a truncated Schechter function (dashed line) using the
  parameters that were derived by a fit to the CMF. The dotted line
  shows the completeness limit used in the fit to the satellite
  contribution.}
\label{f:fig7}
\end{figure*}

In the previous section we derived a model that specifies the stellar
mass of a central galaxy as a function of the virial mass of its host
halo and the stellar mass of a satellite galaxy as a function of the
maximum mass of the subhalo in which it lives. 
It has become common to represent the population of host halos by the
Halo Occupation Distribution (HOD). This includes the halo occupation
function $P(N\vert M)$ which is the probability distribution that a
halo of mass $M$ contains $N$ galaxies (of a specific type).  A close
relative of the HOD is the ``conditional luminosity function''
\citep[CLF; e.g.][]{yang03,vdb2007,yang04}. It extends the halo
occupation function $P(N\vert M)$ (which gives only information about
the total number of galaxies per halo in a given luminosity range)
and yields the average number of galaxies with luminosities in the
range $L \pm {\rm d}L/2$ as a function of the virial mass $M$ of their
host halo.

We define its analog, the ``conditional mass function'' (CMF), or the
average number of galaxies with stellar masses in the range $m \pm
{\rm d}m/2$ as a function of the virial mass $M$ of their host
halo. This provides a direct link between the SMF
$\Phi(m)$ and the host halo mass function ${\rm d}n(M)/{\rm d}M$:
\begin{equation}
\Phi(m) = \int_0^\infty \Phi(m\vert M) \frac{{\rm d}n(M)}{{\rm d}M} \text{d}M
\end{equation}
A host halo of mass $M$ can contain a whole population of galaxies
with different stellar masses $m$. If we count the number of galaxies
living in host halos with a virial mass in the range $M \in [M_1,M_2]$
we can compute the SMF of the halo bin $[M_1,M_2]$:
\begin{equation}
\label{cmftosmf}
\tilde{\Phi}(m) = \int_{M_1}^{M_2} \Phi(m\vert M) \frac{{\rm d}n(M)}{{\rm d}M}
{\rm d}M \approx \Phi(m\vert \bar{M}) \Delta n
\end{equation}
The tilde over a function represents the fact that it is computed in a
halo mass bin. We have replaced the integral by a ``tophat'' with a
width of $\Delta n$ (number of host halos in the bin) and a height of
$\Phi(m\vert M_m)$, where $\bar{M}$ is the geometric mean of the
  minimum and maximum halo masses bracketing the bin.

This equation allows us to put constraints on $\Phi(m\vert M)$ by
calculating $\tilde{\Phi}(m)/\Delta n$. We can then choose an adequate
parameterization of $\Phi(m\vert M)$ and fit these parameters to
$\tilde{\Phi}(m)/\Delta n$ in every halo mass bin. Finally we can
investigate the halo mass dependence of the parameters.

\subsection{Parameterization}
In order to specify the CMF $\Phi(m \vert M)$ we divide the galaxy
population into a central and a satellite part, as in the updated CLF
formalism \citep{zheng2005,zehavi2005,cooray2006,yang08,cacciato2008}.
The central part is $\Phi_c(m \vert M)$ and the satellite part is
$\Phi_s(m \vert M)$. Then the total CMF is the sum of both parts:
\begin{equation}
\Phi(m \vert M) = \Phi_c(m \vert M) + \Phi_s(m \vert M)
\end{equation}

Note that both $\Phi_c(m \vert M)$ and $\Phi_s(m \vert M)$ are
statistical functions and should not be regarded as the mass
functions of galaxies living in a given individual halo.

For the central population we expect the CMF to have a peak around the
stellar mass $m_c$ that corresponds to the host halo's virial mass $M$
in the SHM relation (equation
\ref{eqnmmap}). Due to the halo mass bin size this distribution gets
smeared out, because halos in the interval $[M_1,M_2]$ contain central
galaxies of stellar masses $m \in [m_1(M_1),m_2(M_2)]$. Thus
$\tilde{\Phi}(m)/\Delta n$ will be finite inside the interval
$[m_1(M_1),m_2(M_2)]$ and zero elsewhere with a normalization such
that the number of central galaxies per halo equals one. This can be
regarded as scatter $\sigma_{\rm bin}$ due to the binning. If we add
intrinsic scatter $\sigma_m$ to relation \eqref{eqnmmap}, we expect
$\Phi_c(m\vert M)$ to be a lognormal with a maximum around $m_c(M)$
and a variance of $\sigma_m^2$. To this scatter the binning scatter
$\sigma_{\rm bin}$ adds in quadrature (assuming that $\sigma_{\rm bin}$
and $\sigma_m$ are uncorrelated), resulting in a total scatter of
$\sigma_c^2=\sigma_m^2+\sigma_{\rm bin}^2$. For both cases ($\sigma_m=0$
and $\sigma_m\neq0$) we use a lognormal distribution:
\begin{equation}
\Phi_c(m \vert M) = \frac{1}{\sqrt{2\pi}\ln10\;m\;\sigma_c}
\exp\left[-\frac{\log^2(m/m_c)}{2\sigma_c^2}\right]\;,
\end{equation}
where the mean $m_c(M)$ and width $\sigma_c^2(M)$ are parameterized
functions of the halo mass M.

For the satellite population we adopt a Schechter function with a
steeper slope for the massive end. This is done by squaring the
argument of the exponential function in the Schechter function:
\begin{equation}
\Phi_s(m \vert M) = \frac{\Phi_s^*}{m_s} \left( \frac{m}{m_s}\right)^{\alpha_s}
\exp \left[ -\left( \frac{m}{m_s}\right)^2 \right]\;.
\end{equation}
Also here the parameters $\Phi_s^*(M)$, $m_s(M)$ and $\alpha_s(M)$ are
functions of the host halo mass $M$. They are the normalization, the
characteristic mass and the low mass slope of the satellite population
of host halos of mass $M$.

\subsection{Constraining the conditional mass function} \label{constraincmf}

We populate the halos and subhalos in our simulation with central and
satellite galaxies according to the prescription in section
\ref{s:galaxies}. Then we choose halo mass bins between $\log M/\msun
= 10.2$ and $\log M/\msun = 15.0$ with a bin size of $\Delta M =
0.4{\rm~dex}$. In every halo mass bin we seek all galaxies which live
in a host halo with a mass in that bin, which we divide between
central and satellite galaxies. For these populations we then compute
two seperate SMFs which we normalize
such that the number of central galaxies per host halo equals
one. This procedure then yields for every halo mass bin a central and
a satellite distribution $({\rm d}\tilde{n}_g / {\rm d}\log M) \Delta
n_h$.

Using equation \eqref{cmftosmf} we can now relate the stellar mass
function in a halo mass bin to the CMF:
\begin{eqnarray}
\frac{{\rm d}\tilde{n}_g(m)}{{\rm d}\log M}\frac{1}{\Delta n_h} &=&
\frac{\ln{10}}{\Delta n_h} \; M \; \frac{{\rm d}\tilde{n}_g(m)}{{\rm d}M} \nonumber\\
&=& \ln{10} \; M \; \frac{\tilde{\Phi}(m)}{\Delta n_h} \nonumber\\
&\approx& \ln{10} \; M \; \Phi(m\vert M)
\end{eqnarray}
Now we can fit the five parameters $m_c(M)$, $\sigma_c(M)$, $m_s(M)$,
$\Phi_s^*(M)$ and $\alpha_s(M)$ to the SMFs in each
halo bin. We compute and fit the central and the satellite parts
seperately.

The left panels of Figure \ref{f:fig7} show the CMF in a subsample of
halo mass bins running from $\log M/\msun = 10.2 \pm 0.2 \text{ to }
15.0 \pm 0.2$, where we have not included intrinsic scatter in the
SHM relation. For the satellite part, only galaxies
with a mass above the completness limits for each halo mass bin
(as indicated in Figure \ref{f:fig7}) have been used in the fit.

In low-mass halos ($\log M/\msun<11.0$) the contribution from
satellite galaxies is very small and the central contribution
dominates until $\log M/\msun=12.0$. For massive halos ($\log
M/\msun>13.0$) the satellite contibution dominates by number. The mean
of the lognormal fit to the central contribution also increases with
halo mass as stipulated by the model derived in
Section~\ref{s:galaxies}. The characteristic mass scale of the
satellite contribution also increases with halo mass meaning that the
most massive satellite galaxies have a mass which is comparable to the
mass of the central galaxy.

The scatter of the central contribution $\sigma_c(M)$ decreases with
halo mass. As we did not include any scatter in the model, this
scatter reflects the width ($0.4{\rm~dex}$) of the halo mass bins
($\sigma_{\rm bin}$). The halo mass dependence of $\sigma_c(M)$
arises because a fixed halo mass bin is mapped to a smaller galaxy mass bin
for larger halo mass due to the shape of the SHM relation.
Another feature of the CMF is the slope for low mass satellite galaxies
$\alpha_s(M)$ which becomes shallower with increasing halo mass.

\subsection{The parameters of the conditional mass function}\label{s:cmfpars}

In this section we investigate the halo mass dependence of the five
parameters of the CMF: $m_c(M)$, $\sigma_c(M)$, $m_s(M)$,
$\Phi_s^*(M)$ and $\alpha_s(M)$. They have been fixed by fitting to
the stellar mass functions in each halo mass bin. We introduce a
parameterization in order to describe the dependence on halo mass and
constrain these by a fit to each parameter. The results are presented
in Table \ref{t:cmfparameters}. This provides a complete description
of the CMF.

\begin{figure*}
\centering
\epsscale{1.15}
\plotone{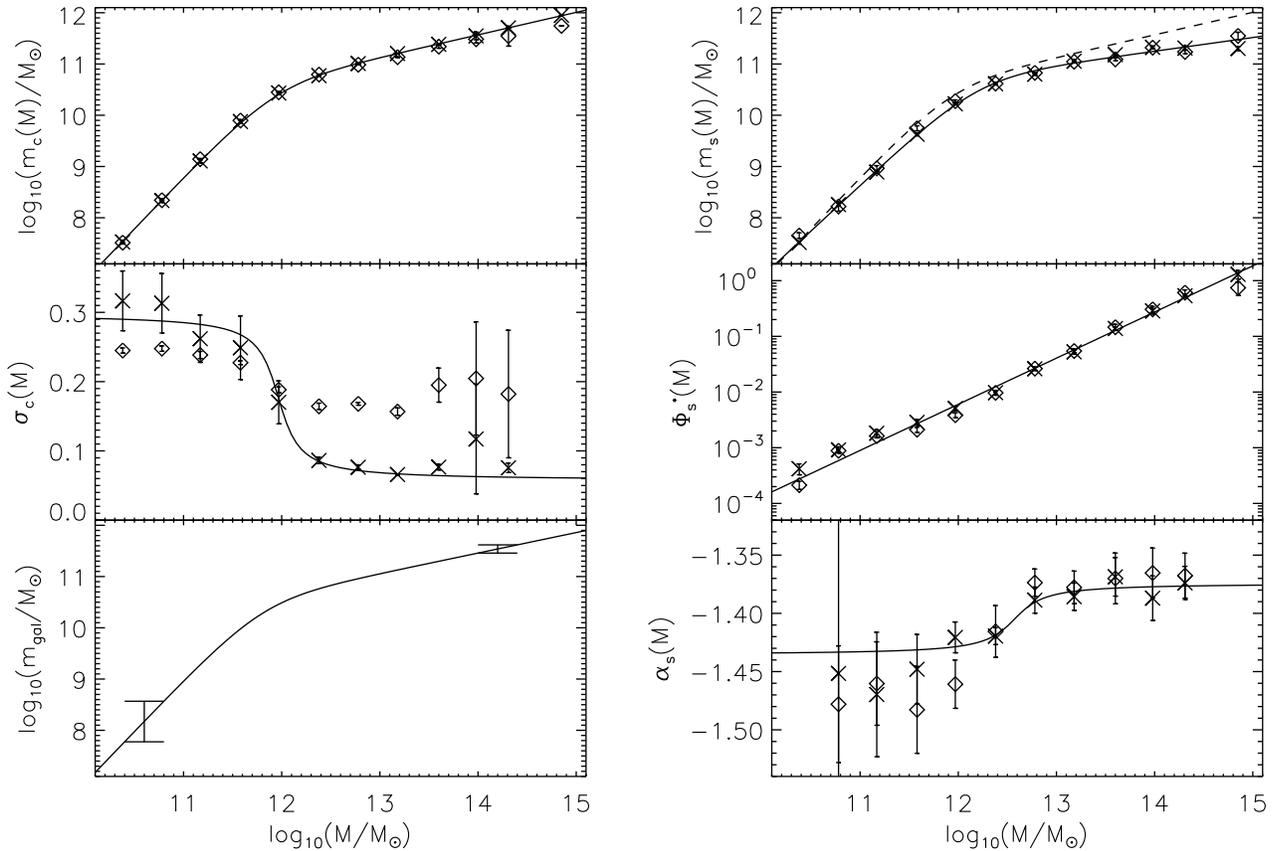}
\caption{The five parameters of the conditional mass function as a
  function of halo mass. The crosses were derived from a fit to the
  CMF in every halo mass bin (assuming no scatter in stellar-to-halo
  mass relation). The solid line is a fit to the crosses using the
  respective parameterization. The CMF parameters derived with a
  scatter of $\sigma_m=0.15$ in the stellar-to-halo mass relation are
  given by the diamonds. The left panels show the central
  contribution: $m_c(M)$ (top), $\sigma_c(M)$ (middle) and an
  illustration of the behavior of $\sigma_c(M)$ (bottom). The right
  panels show the satellite contribution: $m_s(M)$ (top),
  $\Phi_s^*(M)$ (middle) and $\alpha_s(M)$ (bottom). The dashed line
  in the top right panel indicating $m_c(M)$ has been added for
  comparison}
\label{f:fig8}
\end{figure*}

\begin{deluxetable}{lllll}
\tablecaption{Parameters of the CMF} \tablehead{ \colhead{} &
  \colhead{$\sigma_m=0.0$} & \colhead{} & \colhead{$\sigma_m=0.15$} &
  \colhead{} } \startdata $\log M_{1c}$ & 11.9347 & $\pm$ 0.0257 &
11.9008 & $\pm$ 0.0119 \\ $(m_c/M)_0$ & ~~0.0267 & $\pm$ 0.0006 &
~~0.0297& $\pm$ 0.0004 \\ $\beta_c$ & ~~1.0059 & $\pm$ 0.0332 &
~~1.0757& $\pm$ 0.0097 \\ $\gamma_c$ & ~~0.5611 & $\pm$ 0.0065 &
~~0.6310 & $\pm$ 0.0121 \\ \tableline $\log M_2$ & 11.9652 & $\pm$
0.1118 & 11.8045 & $\pm$ 0.0458 \\ $\sigma_{\infty}$ & ~~0.0569 &
$\pm$ 0.0052 & ~~0.1592& $\pm$ 0.0030 \\ $\sigma_1$ & ~~0.1204 & $\pm$
0.0191 & ~~0.0460& $\pm$ 0.0029 \\ $\xi$ & ~~6.3020 & $\pm$ 3.0720 &
~~4.2503& $\pm$ 0.9945 \\ \tableline $\log M_{1s}$ & 12.1988 & $\pm$
0.0878 & 12.0640& $\pm$ 0.0931 \\ $(m_s/M)_0$ & ~~0.0186 & $\pm$
0.0012 & ~~0.0198& $\pm$ 0.0015 \\ $\beta_s$ & ~~0.7817 & $\pm$ 0.0629
& ~~0.8097 & $\pm$ 0.0971 \\ $\gamma_s$ & ~~0.7334 & $\pm$ 0.0452 &
~~0.6910& $\pm$ 0.0390 \\ \tableline $-\log\Phi_0$ & 11.1622 & $\pm$
0.2874 & 10.8924& $\pm$ 0.4615 \\ $\lambda$ & ~~0.8285 & $\pm$ 0.0215
& ~~0.8032 & $\pm$ 0.0367 \\ \tableline $\log M_3$ & 12.5730 & $\pm$
0.1351 & 12.3646 & $\pm$ 0.0260 \\ $-\alpha_{\infty}$ & ~~1.3740 &
$\pm$ 0.0066 & ~~1.3676 & $\pm$ 0.0043 \\ $-\alpha_1$ & ~~0.0309 &
$\pm$ 0.0076 & ~~0.0524& $\pm$ 0.0051 \\ $\zeta$ & ~~4.3629 & $\pm$
2.6810 & ~~9.5727 & $\pm$ 6.8240 \\ \enddata 
\tablecomments{The second
  and third columns give the CMF parameters and their errors for a
  model without scatter while the fourth and the fifth columns give
  the CMF parameters and their errors for a model with a scatter of
  $\sigma_m=0.15$. All quoted masses are in units of
  $\msun$ \vspace{0.3cm}}
\label{t:cmfparameters}
\end{deluxetable}

As we have already determined the mean relation between the stellar
mass of a galaxy and the mass of its halo, the form of $m_c(M)$ has to
be the same and can thus be decribed by equation \eqref{eqnmmap}:
\begin{equation}
\label{mcmap}
m_c(M) = 2 \; M \; \left( \frac{m_c}{M}\right)_0 \left[\left(\frac{M}{M_{1c}}\right)^{-\beta_c} + \left(\frac{M}{M_{1c}}\right)^{\gamma_c}\right]^{-1}
\end{equation}
This yields four parameters $(m_c/M)_0$, $M_{1c}$, $\beta_c$ and $\gamma_c$.

In the upper left panel of Figure \ref{f:fig8} $m_c(M)$ is plotted as
a function of halo mass. Note that by construction, it has the same form
as the SHM relation.

\label{sigmac}
The scatter of the central galaxy contribution is high for low halo
masses and decreases for more massive halos. The middle left panel of
Figure~\ref{f:fig8} shows $\sigma_c(M)$ as a function of halo mass. As
one can see, $\sigma_c(M)$ goes to a constant value both for low and
high halo masses while it decreases with halo mass. We therefore
choose the following parameterization:

\begin{equation}
\label{scmap}
\sigma_c(M) = \sigma_{\infty} + \sigma_1 \left[ 1-\frac{2}{\pi} \arctan\left( \xi \log \frac{M}{M_2} \right) \right]
\end{equation}
This yields four more parameters $\sigma_{\infty}$, $\sigma_1$, $\xi$
and $M_2$. Here, $\sigma_{\infty}$ sets the high mass limit of
$\sigma_c(M)$ while $\sigma_1$ sets the difference between the low and
high mass limits of $\sigma_c(M)$. The parameter $M_2$ determines the
mass scale at which the transition occurs and $\xi$ sets the
strength. For a large (small) value of $\xi$ the transition occurs in
a small (large) interval around $M_2$.

The specific shape of $\sigma_c(M)$ can be explained by the form of
the SHM relation (equation \ref{eqnmmap}). As we have not included any
scatter in this relation ($\sigma_m=0$), the width of the lognormal
function of the central galaxy distribution arises from the width of
the halo mass bin ($\sigma_c=\sigma_{\rm bin}$). A halo mass interval
$[M_1,M_2]$ contains only central galaxies with stellar masses of $m
\in [m_1(M_1),m_2(M_2)]$. The lower left panel of Figure~\ref{f:fig8}
illustrates this by showing how halo mass bins affect the bin size of
the stellar mass. If we choose the same bin size for low and high mass
halos, we get different bin sizes for low and high mass galaxies, due
to the changing slope of $m(M)$. Therefore the transition occurs where
the slope of $m(M)$ changes which is around $M_1$, so the value of
$M_2$ is very close to that value.

As Figure~\ref{f:fig7} shows that the satellite contribution falls off
around the mean mass of the central galaxy, we expect the
characteristic mass of the modified Schechter function $m_s(M)$ to
follow $m_c(M)$. We therefore describe $m_s(M)$ with the same function
we used for the parametrisation of $m_c(M)$:
\begin{equation}
\label{msmap}
m_s(M) = 2 \; M \; \left( \frac{m_s}{M}\right)_0 \left[\left(\frac{M}{M_{1s}}\right)^{-\beta_s} + \left(\frac{M}{M_{1s}}\right)^{\gamma_s}\right]^{-1}
\end{equation}
This function yields four parameters $(m_s/M)_0$, $M_{1s}$, $\beta_s$
and $\gamma_s$.

The upper right panel of Figure~\ref{f:fig8} plots $m_s(M)$ as a
function of halo mass.
We see that the shape is similar to that of $m_c(M)$.
Note that $m_s(M)$ is always lower than $m_c(M)$, while the deviation
increases with increasing halo mass. This implies that for high
halo masses the satellite contribution to the CMF falls off before the
mean mass of the central galaxy.

The normalization of the modified Schechter function is small for low
halo masses and increases with the mass of the host halo. The middle
right panel of Figure \ref{f:fig8} shows $\Phi_s^*(M)$ as a function of
halo mass. We see that $\Phi_s^*(M)$ can be described by a power law and
choose the following parametrisation:
\begin{equation}
\label{psmap}
\Phi_s^*(M) = \Phi_0 \left( \frac{M}{\msun}\right)^{\lambda} 
\end{equation}
We get two more parameters, $\Phi_0$ and $\lambda$. The normalization
of $\Phi_s^*(M)$ is given by $\Phi_0$ and the slope by $\lambda$.
The shape of $\Phi_s^*(M)$ implies that the probability for a host halo to
harbor satellite galaxies (in a given stellar mass range) increases with
increasing halo mass.

\label{alphas}
The slope of the modified Schechter function for the satellite
contribution becomes shallower for more massive
halos. The lower right panel of Figure~\ref{f:fig8} shows
$\alpha_s(M)$ as a function of halo mass and shows that $\alpha_s(M)$
goes to a constant value for both low and high halo masses.
Similar to $\sigma_c(M)$, we choose the parameterization:
\begin{equation}
\label{asmap}
\alpha_s(M) = \alpha_{\infty} + \alpha_1 \left[ 1-\frac{2}{\pi} \arctan\left( \zeta \log \frac{M}{M_3} \right) \right]
\end{equation}
This yields four more parameters $\alpha_{\infty}$, $\alpha_1$,
$\zeta$ and $M_3$. Here, $\alpha_{\infty}$ sets the high mass limit of
$\alpha_c(M)$ while $\alpha_1$ sets the difference between the low and
high mass limits of $\alpha_c(M)$. The mass scale at which this
transition occurs is determined by $M_3$ and $\zeta$ sets its
strength. The transition occurs in a small (large) interval around
$M_3$ for a large (small) value of $\zeta$.

\begin{figure*}
\centering
\epsscale{1.15}
\plotone{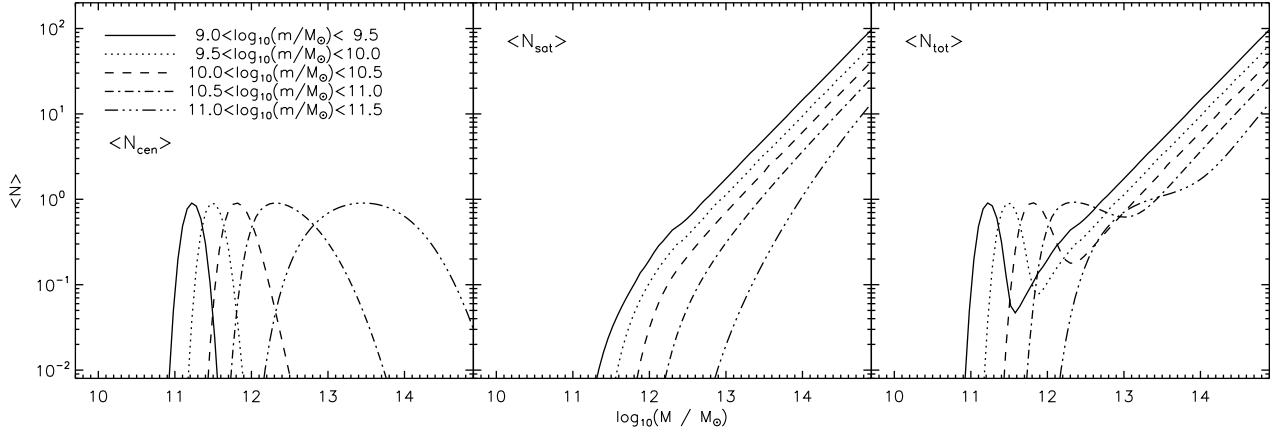}
\caption{Occupation numbers as function of halo mass in stellar mass
  bins, derived using the conditional mass function. The left, middle
  and right panels show the average number of central, satellite and
  total galaxies per halo, respectively.}
\label{f:fig9}
\end{figure*}

\subsection{The impact of scatter}

Until now, we have used the SHM relation \eqref{eqnmmap} without any
intrinsic scatter. In this section we investigate how the CMF and the
parameters change if we include a scatter $\sigma_m$ as described in
section \ref{s:scattersec}. This scatter is again assumed to be
constant with host halo mass.

The right panels of Figure~\ref{f:fig7} show the resulting CMF in a
subsample of halo mass bins for an intrinsic scatter of $\sigma_m =
0.15$. The central part is now no longer near-constant in the interval
$[m(M-\Delta M/2),m(M+\Delta M/2)]$ as in the left panels of
Figure~\ref{f:fig7} (where $\sigma_m = 0.0$) but has the form of a
lognormal with a broader distribution for bigger $\sigma_m$. As the
scatter has been taken from a lognormal distribution, the central
galaxy contribution to the CMF is distributed in the same way. Hence,
$\sigma_c(M)$ changes with respect to the model that does not include
artificial scatter. We notice that at the massive end the binning
scatter $\sigma_{\rm bin}^2$ and the intrinsic scatter $\sigma_m^2$ add to
the total scatter $\sigma_{\rm tot}^2$.
At the low mass end, however, the total scatter is less than what has
been obtained by using no intrinsic scatter. This shows that the two forms
of scatter do not add in quadrature and indicates that they are correlated.

We compare $m_c(M)$, $\sigma_c(M)$, $m_s(M)$, $\Phi_s^*(M)$ and
$\alpha_s(M)$ for $\sigma_m=0$ and $\sigma_m=0.15$ and show the
resulting parameters in Table \ref{t:cmfparameters} (columns four and
five) and in Figure~\ref{f:fig8}. The mean mass of the central galaxy
$m_c(M)$ does not change much if artificial scatter is introduced. The
most likely stellar mass of a central galaxy is still given by the
SHM relation, so the mean of the gaussian in
logarithmic space stays the same. Also the parameters of the satellite
population [$m_s(M)$,$\Phi_s^*(M)$ and $\alpha_s(M)$] do not change
significantly.

\subsection{The occupation numbers}

\begin{figure*}
\centering
\epsscale{1.15}
\plotone{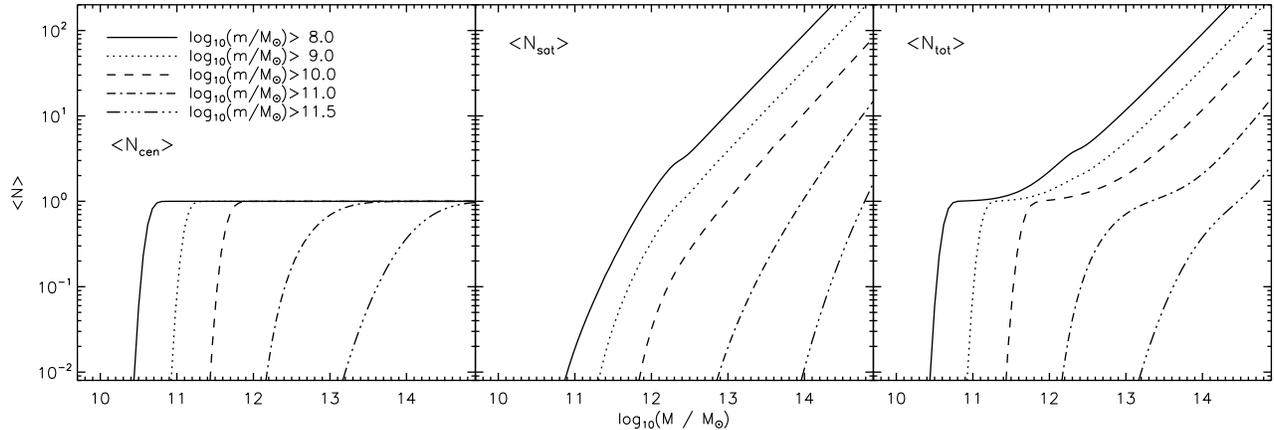}
\caption{Occupation numbers as function of halo mass for galaxy
  samples with a stellar mass above a given threshold. The left,
  middle and right panels show the average number of central,
  satellite and total galaxies per halo, respectively.}
\label{f:fig10}
\end{figure*}

In order to compare our results to other HOD models it is useful to
compute the average number of galaxies per halo $\langle N\rangle$, as
this is the main prediction of the HOD approach. To compute $\langle
N\rangle(M)$ from the CMF we simply integrate $\Phi(m\vert M)$ over
the desired stellar mass range:
\begin{equation}
\langle N\rangle(M) = \int_{m_1}^{m_2} \Phi(m\vert M) {\rm d}m
\end{equation}
As we have divided $\Phi(m\vert M)$ into a central galaxy contribution
$\Phi_c(m\vert M)$ and a satellite galaxy contribution $\Phi_s(m\vert
M)$, we can compute seperate occupation numbers for central and
satellite galaxies:
\begin{eqnarray*}
\langle N\rangle(M) &=& \int_{m_1}^{m_2} \Phi_c(m\vert M) {\rm d}m + \int_{m_1}^{m_2} \Phi_s(m\vert M) {\rm d}m\notag\\
&=& \langle N_c\rangle(M) + \langle N_s\rangle(M)
\end{eqnarray*}
The average number of central galaxies per halo $\langle
N_c\rangle(M)$ is given by
\begin{equation}
\label{eqnocccen}
\langle N_c\rangle(M) = \frac{1}{2}\left[{\rm erf}(\eta_2)-{\rm erf}(\eta_1)\right] \;,
\end{equation}
with the error-function ${\rm erf}(x)$ and the integration boundaries
\begin{equation*}
\eta_1 = \frac{\log(m_1/m_c)}{\sqrt{2}\sigma_c} \quad \text{and} \quad \eta_2 = \frac{\log(m_2/m_c)}{\sqrt{2}\sigma_c} \;.
\end{equation*}
The average number of satellite galaxies per halo $\langle N_s\rangle(M)$ is
\begin{equation}
\label{eqnoccsat}
\langle N_s\rangle(M) = \frac{\Phi_s}{2}\left[\Gamma\left(\frac{\alpha_s}{2}+\frac{1}{2},\kappa_1\right)-\Gamma\left(\frac{\alpha_s}{2}+\frac{1}{2},\kappa_2\right)\right] \;,
\end{equation}
with the upper incomplete gamma function $\Gamma(a,x)$ and the
integration boundaries
\begin{equation*}
\kappa_1 = (m_1/m_s)^2 \quad \text{and} \quad \kappa_2 = (m_2/m_s)^2 \;.
\end{equation*}

Figure~\ref{f:fig9} shows the resulting occupation numbers for the
values of the CMF parameters that were derived in section
\ref{s:cmfpars} (using a scatter of $\sigma_m=0.15$).
The five lines in each panel correspond to different stellar mass bins.

The left panel shows the average number of central galaxies per halo
$\langle N_c\rangle(M)$ as a function of halo mass. In the middle
panel, the average number of satellite galaxies per halo $\langle
N_s\rangle(M)$ as a function of halo mass is shown. The right panel
plots the average number of all galaxies per halo $\langle
N_{\rm tot}\rangle(M)$ as a function of halo mass. A galaxy of a low
stellar mass can thus either be a central galaxy of a low mass halo,
or a satellite galaxy of a massive halo. It is not likely to live in a
halo of intermediate mass.

As it is common in the literature to plot occupation numbers not for
stellar mass intervals, but for galaxy samples with a mass above a
given threshold, we need to adjust equations \eqref{eqnocccen} and
\eqref{eqnoccsat}. The stellar mass threshold is then given by $m_1$
while $m_2 \to \infty$. This yields for the average number of central
galaxies
\begin{equation}
\langle N_c\rangle(M, m_1) = \frac{1}{2}\left[1 - {\rm erf}\left(\frac{\log(m_1/m_c)}{\sqrt{2}\sigma_c}\right)\right]\;,
\end{equation}
since ${\rm erf}(x\to\infty)\to 1$, and for the average number of satellite galaxies
\begin{equation}
\langle N_s\rangle(M, m_1) = \frac{\Phi_s}{2} \; \Gamma\left[\frac{\alpha_s}{2}+\frac{1}{2},\left( \frac{m_1}{m_s}\right)^2\right]\;
\end{equation}
since $\Gamma(a,x\to\infty)\to 0$.

Figure \ref{f:fig10} shows occupation numbers for different stellar
mass thresholds. The left panel shows the average number of central
galaxies per halo $\langle N_c\rangle(M)$ as a function of halo
mass. The middle panel plots the average number of satellite galaxies
per halo $\langle N_s\rangle(M)$ as a function of halo mass. It is
similar to the middle panel of Figure~\ref{f:fig9} while it is larger
at a given halo mass. In the right panel the average number of all
galaxies per halo $\langle N_{tot}\rangle(M)$ as a function of halo
mass is shown.

\section{Comparison}\label{s:comparison}

\subsection{Other HOD models}

\begin{deluxetable}{lllll}
\tablecaption{Comparison between different models} \tablehead{
  \colhead{} & \colhead{$\log M_1$} & \colhead{$(m/M)_0$} &
  \colhead{$\beta$} & \colhead{$\gamma$} } \startdata Our model &
11.884 & 0.0282 & 1.06 & 0.556\\ Non-Parametric & 11.766 & 0.0324 &
1.43 & 0.565\\ \citet{wang06} & 11.845 & 0.0319 & 1.42 &
0.710\\ Somerville SAM & 11.888 & 0.0276 & 0.98 & 0.629\\ Croton SAM &
11.742 & 0.0405 & 0.92 & 0.610\\ Yang GC & 12.067 & 0.0384 &
0.71 & 0.698\\
\enddata
\tablecomments{All quoted masses are in units of $\msun$}
\label{t:paracomparison}
\end{deluxetable}

Numerous variations on halo occupation models have been presented in
the literature. In this section we describe some of the most popular
ones and compare them to our model. As many authors use different
initial mass functions and definitions of halo masses, we convert all
results to the conventions that we have used in this work (Kroupa IMF
and virial overdensity).

In the Non-Parametric model \citep{valeostriker06,conroy06,shankar2006},
galaxy properties, such as luminosity and stellar mass, are monotonically
related to the mass of dark matter halos.
Using the observed galaxy SMF, the most massive halo
is matched to the most massive galaxy:
\begin{equation}
n_g(>m_i) = n_h(>M_i)
\end{equation}
In this way, the observed SMF is automatically
reproduced. Applying this procedure and fitting the parameters of the
SHM relation to the result, we have derived the
values given in Table \ref{t:paracomparison}. These are in good
agreement with the parameters of our model, except for $\beta$. We
find that this is due to the shape of the SHM
ratio for low masses. For the Non-Parametric model, $m(M<M_1)$ can not
be perfectly described by a single power law, as is assumed in our
model. 

Adding an additional parameter and assuming a fitting function with
five free parameters, we are able to fit the SHM relation predicted by
the non-parametric model quite precisely. The fifth parameter accounts
for the deviation from the power-law at high and low masses. Using the
parameterization
\begin{equation}
\label{eqnnonpar}
m(M)=m_0 \; \frac{(M/M_1)^{\gamma_1}}{\left[1+(M/M_1)^{\beta}\right]^{(\gamma_1-\gamma_2)/\beta}}
\end{equation}
we determine the values given in Table
\ref{t:paranonpar}. Figure~\ref{f:fig11} shows the results of four-
and five-parameter fits to the SHM relation derived via the
non-parametric method, compared with our usual model. In the range
where we applied the mass function fit, the non-parametric model lies
within our error-bars.

\begin{deluxetable}{llllll}
\tablecaption{Fit parameters for Equation~\eqref{eqnnonpar}}
\tablehead{
  \colhead{ } &
  \colhead{$\log m_0$} &
  \colhead{$\log M_1$} &
  \colhead{$\gamma_1$} &
  \colhead{$\gamma_2$} &
  \colhead{$\beta$}
}
\startdata
 & 10.864 & 10.456 & 7.17 & 0.201 & 0.557\\
$\pm$ & ~~0.043 & ~~0.211 & 1.16 & 0.018 & 0.031\\

\enddata
\tablecomments{All masses are in units of $\msun$}
\label{t:paranonpar}
\end{deluxetable}

In \citet{wang06} a model similar to ours is used to constrain the
SHM ratio. The halo catalogue is taken from the
Millennium simulation \citep{springel05b}; halos are identified using
a friends-of-friends group finder while substructure is found using
the {\small SUBFIND} algorithm of \citet{springel01}. As observational
constraints, the authors use a SMF which they
compute from the SDSS DR2 data using the mass estimates of
\citet{kauffmann03} and the projected CFs of
\citet{li06}.

The parameterization they use is similar to ours, with four free
parameters that can easily be converted to $M_1$, $(m/M)_0$, $\beta$
and $\gamma$ and an unconstrained scatter. These are fixed by
generating a grid of models and the best-fit model is defined as the
one for which $\chi^2=\chi^2(\Phi)+\chi^2(w_p)$ is minimal. They find
that their fit improves if they take a different set of parameters for
central and satellite galaxies. In Table~\ref{t:paracomparison} we
compare our best-fit parameters with their central galaxy best-fit
parameters which have been updated in \citet{wang07}. We show these
results in Figure~\ref{f:fig11}.

The values of $M_1$ and $(m/M)_0$ are in very good agreement with our
values, but the slopes are both higher, resulting in fewer massive and
fewer low mass galaxies. The reason for the difference in the low mass
end is the different simulation used. As the resolution of the
simulation in our model is higher, the low mass end can be constrained
more tightly. For the massive end the difference in $\gamma$ can be
explained by the additional unconstrained scatter that is used in
\citet{wang06}. As the mass function is steep at high masses and
shallow for low masses, a change in the scatter will influence the
number of massive galaxies strongly, while it will have only a small
effect on the low mass end. As the other three parameters $M_1$,
$(m/M)_0$ and $\beta$ are coupled to the Schechter function
parameters, there are two parameters to constrain the slope of the
massive end of the SMF. This degeneracy can cause
the difference in $\gamma$ between the two models. The fact that in
the Millennium simulation the cosmology is different to that of our
simulation also affects the value of the parameters.

\subsection{Gravitational lensing}

\begin{figure}
\centering
\epsscale{1.2}
\plotone{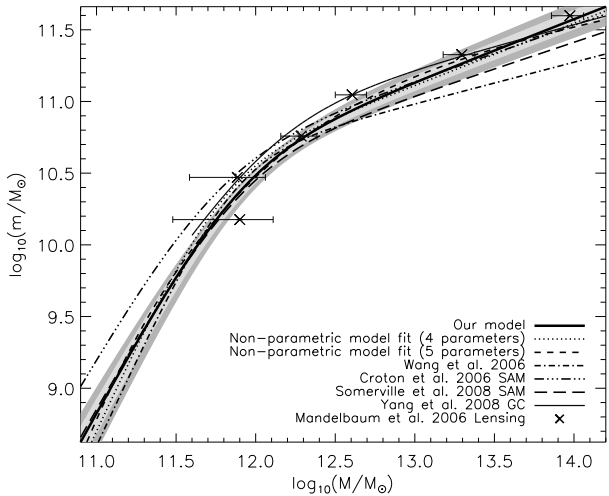}
\caption{Comparison of the stellar-to-halo mass relation $m(M)$
  between our model (solid line), models from other authors and
  galaxy-galaxy lensing (symbols). The blue areas are the 1-$\sigma$
  and 2-$\sigma$ levels and the error-bars on the symbols are the
  2-$\sigma$ levels of the halo mass.}
\label{f:fig11}
\end{figure}

The relation between stellar mass and halo mass can be constrained
observationally using galaxy-galaxy lensing. Gravitational lensing
induces shear distortions of background objects around foreground
galaxies, allowing the mass of the dark matter halo to be
estimated. \citet{mandelbaum05,mandelbaum06} have used
SDSS data to calibrate the predicted signal from a halo model which
has been derived from a dissipationless simulation. They have
extracted the mean halo mass as a function of stellar mass. The
lensing data for combined early and late-type galaxies
(Mandelbaum, private communication)
are shown in Figure~\ref{f:fig11} and are in
excellent agreement with our model.

\subsection{Semi-analytic models}
\label{sams}

As we discussed in the introduction, semi-analytic models (SAMs) of
galaxy formation attempt to predict the relationship between dark halo
mass and stellar mass by a priori modelling of physical processes,
such as the growth of structure, cooling, star formation, and stellar
and AGN feedback. We compare our results with predictions from the
latest version of the semi-analytic models of \citet{somerville1999};
see \citet{somerville2008}. For this we compute the mean stellar mass
of central galaxies as a function of the mass of the host halo in halo
mass bins. The results are shown in Figure~\ref{f:fig11} and are in
good agreement with our model. This is not surprising, as the physical
parameters in the model of \citet{somerville2008} have been tuned to
match the observed stellar mass function at $z=0$.

In \citet{wang06} the authors use the semi-analytic model of
\citet{croton06} and link galaxy properties, such as the stellar mass,
to the mass of the halo in which the galaxy was last a central object
$M_{\rm infall}$. They fit the same four-parameter function that they
used for their empirical model (described above) to obtain the
parameter estimates from the SAM. We summarize these results in
Table~\ref{t:paracomparison}, and show them in Figure~\ref{f:fig11}.

The two slopes are in very good agreement with our results. However,
the normalization in the \citet{croton06} SAM is $\sim25\%$ higher and
the characteristic mass is $\sim25\%$ lower than what we found and
what \citet{wang06} find for their model. This is because the SAM of
\citet{croton06} does not produce a perfect fit to the observed
SMF.

\subsection{SDSS group catalogue}
\label{groupcat}

Another direct way of studying galaxy properties as a function of halo
mass is using the SDSS group catalogue presented in \citet{yang07}. In
this approach, galaxies are first linked together into ``groups''
using a friends-of-friends algorithm. Each group is then assigned a
total halo mass by matching to the theoretical dark matter halo mass
function. \citet{yang08} present the relation between the mean stellar
mass of the central galaxy and the host halo mass. We fit the
parameters of equation \ref{eqnmmap} to their relation and present the
results in Table~\ref{t:paracomparison}.

We note that the characteristic mass and the normalization derived
from the group catalogue are both higher than our model
parameters. The high mass slope of the SHM relation in the group
catalogue is shallower than that of our model. The low mass slope is
also shallower, however, the constraints on the low mass slope in the
group catalogue are weak, since the lowest halo masses are
$log(M/\msun)\sim11.7$.  This can also be seen in Figure~\ref{f:fig11}
where we show the SHM relation of the group catalogue for comparison.

\section{High Redshift} \label{s:redshift}

The discussion in the previous sections has focussed solely on the
present day universe. In this section we extend our analysis to higher
redshifts and derive the redshift dependence of the stellar-to-halo
mass relation. Having chosen a particular observed stellar mass
function at a given redshift, we can investigate how the parameters of
the SHM ratio change with time. This allows us to
learn about the evolution of galaxies.  Also, with this information,
we can populate the $N$-body simulation snapshots with galaxies at
different redshifts using the appropriate redshift dependent
SHM relation, and then use the spatial information
from the simulation to compute the stellar mass dependent correlation
functions.

Since at the present time there are no high redshift ($z\gta 1$)
clustering data as a function of stellar mass available, we fit the
four parameters of equation \eqref{eqnmmap} to the observed SMFs at
a given redshift. We argued in section \ref{combinedfit} that, under
the assumption that central and satellite galaxies follow the same
SHM relation, the SMFs provide much stronger constraints on the SHM
ratio than the clustering data. Thus we should be able to use our
model to predict clustering as a function of stellar mass at any
redshift.

\begin{figure*}
\centering
\epsscale{1.1}
\plotone{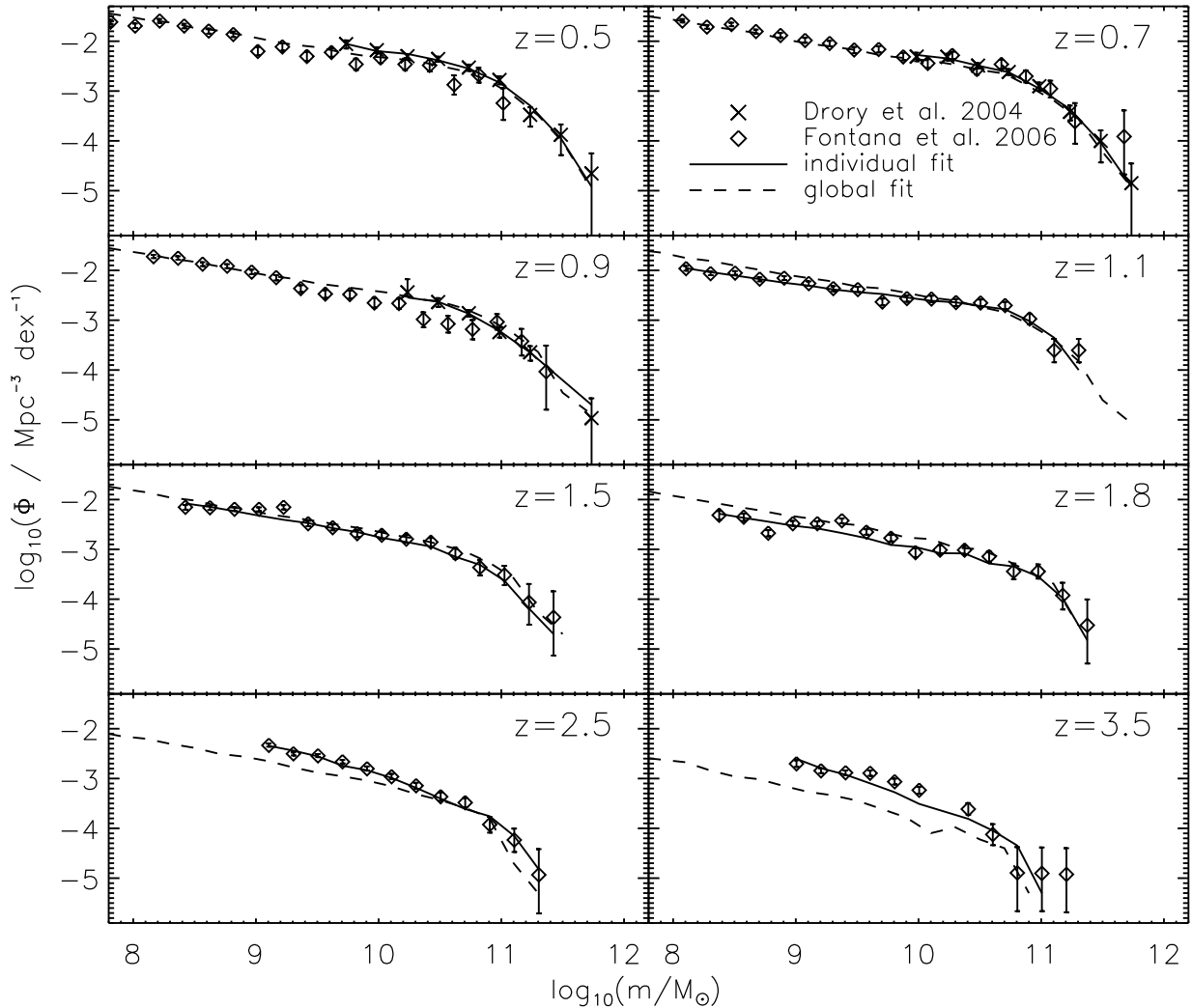}
\caption{Comparison between the model and the observed stellar mass
  functions for different redshifts. The observed stellar mass
  functions are taken from \citet{drory2004} (for $z\leqslant 0.9$)
  and from \citet{fontana2006} (for $z\geqslant 1.1$) and are
  represented by the symbols. The model stellar mass functions have
  been fitted to the observations and are represented by the solid
  lines. The dashed lines are the theoretical mass function we obtain
  from the redshift-dependent parameterization. The redshift is
  indicated at the top of each panel.}
\label{f:fig12}
\end{figure*}

\begin{figure*}
\centering
\epsscale{1.15}
\plotone{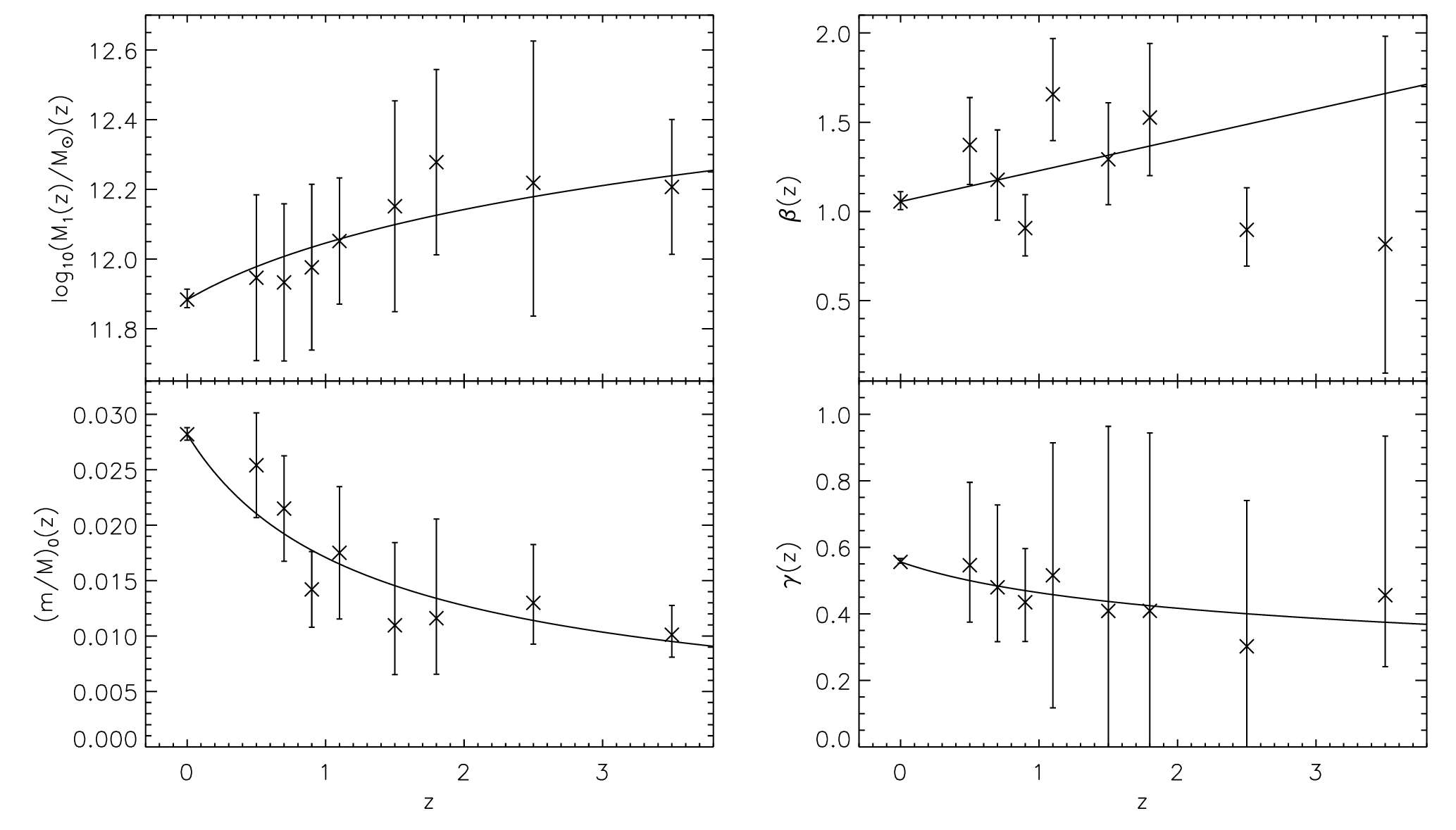}
\caption{Evolution of the stellar-to-halo mass relation parameters
  with redshift. The symbols correspond to the derived values while
  the solid line is a fit to the data. For $M_1$, $(m/M)_0$ and
  $\gamma$ this is a power-law, while for $\beta$ it is a straight
  line.}
\label{f:fig13}
\end{figure*}

\subsection{Which survey for which redshift}

In order to constrain the SHM relation we have to first select
observational stellar mass functions at the redshifts we want to
investigate. Because of the trade-off between surveying large areas
and obtaining deep samples, measurements of the SMF at high redshift
tend to suffer from limited dynamic range. Therefore it is important
to think about how the constraints on our four SHM function parameters
arise from the observations.

The characteristic mass $M_1$ and the maximum SHM
ratio $(m/M)_0$ mostly depend on galaxies and halos of intermediate
mass. The high mass slope $\gamma$ is fixed by the number of massive
galaxies since these live in the massive halos. On the other hand, the
low mass slope $\beta$ is set by the number of low mass galaxies since
these live in the low mass halos.

For a survey with a fixed area on the sky, the observed volume is
smaller for low redshifts $(z\lesssim1)$ than for high redshifts.  In
order to compute the SMF at high galaxy masses, the
observed volume has to be relatively large, as massive galaxies are
rare. Thus for low redshifts one has to choose a wide survey (large
area) to determine the SMF for massive galaxies and
properly constrain $\gamma$. Constraining the SMF at
the low mass end requires a high level of completeness for low mass
galaxies, which are very faint objects. Hence we have to choose a deep
survey that can detect faint galaxies in order to constrain $\beta$.

Taking these considerations into account, we choose the stellar mass
functions presented in \citet{drory2004} to constrain the parameters
$M_1$, $(m/M)_0$ and $\gamma$ at low redshifts. The authors derive the
SMFs using MUNICS which is a wide area, medium-deep
survey selected in the $K$ band. The detection limit is $K\approx19.5$
and the subsample the authors use covers $0.28 {\rm~deg}^2$. We apply
our method using these mass functions and take the three parameters
from that analysis.

However, the MUNICS survey is not deep enough to detect galaxies that
are fainter than the characteristic mass of the SMF (the knee) and
thus is not sufficient to constrain the parameter $\beta$.  To
constrain $\beta$ we choose the SMFs derived in
\citet{fontana2006}. This work is based on the GOODS-MUSIC sample, a
multicolor catalogue extracted from the survey conducted over the
Chandra Deep Field South. The catalogue is selected in the $z_{850}$
and $K$ bands, covers an area of $143.2 {\rm~arcmin}^2$, and is
complete to a typical magnitude of $K\approx23.5$. We apply our method
using the SMFs computed with the $z_{850}$ band selected sample and
take the parameter $\beta$ from that analysis.

For high redshift $(z\gtrsim1)$ we use the SMFs presented in
\citet{fontana2006} to constrain all four parameters. For high
redshifts, the volume of a redshift bin becomes large enough to sample
massive galaxies, and therefore the GOODS-MUSIC sample is sufficient
to constrain $\gamma$.

We convert all SMFs which use a Salpeter initial mass function to the
Kroupa/Chabrier initial mass function.

\subsection{Evolution of the parameters} \label{s:parevo}

\begin{deluxetable}{c cc cc ccc cc}
\tablecaption{\small Stellar-to-halo mass ratio parameters for different redshifts}
\tablehead{
  \colhead{z} &
  \colhead{\scriptsize $\log M_1$} &
  \colhead{\scriptsize $\pm$} &
  \colhead{\scriptsize $(m/M)_0$} &
  \colhead{\scriptsize $\pm$} &
  \colhead{\scriptsize $\beta$} &
  \colhead{\scriptsize $-$} &
  \colhead{\scriptsize $+$} &
  \colhead{\scriptsize $\gamma$} &
  \colhead{\scriptsize $\pm$}
}
\startdata
\scriptsize 0.0 & \scriptsize 11.88 & \scriptsize 0.02 & \scriptsize 0.0282 & \scriptsize 0.0005 & \scriptsize 1.06 & \scriptsize 0.05 & \scriptsize 0.05 & \scriptsize 0.56 & \scriptsize 0.00\\
\scriptsize 0.5 & \scriptsize 11.95 & \scriptsize 0.24 & \scriptsize 0.0254 & \scriptsize 0.0047 & \scriptsize 1.37 & \scriptsize 0.22 & \scriptsize 0.27 & \scriptsize 0.55 & \scriptsize 0.17\\
\scriptsize 0.7 & \scriptsize 11.93 & \scriptsize 0.23 & \scriptsize 0.0215 & \scriptsize 0.0048 & \scriptsize 1.18 & \scriptsize 0.23 & \scriptsize 0.28 & \scriptsize 0.48 & \scriptsize 0.16\\
\scriptsize 0.9 & \scriptsize 11.98 & \scriptsize 0.24 & \scriptsize 0.0142 & \scriptsize 0.0034 & \scriptsize 0.91 & \scriptsize 0.16 & \scriptsize 0.19 & \scriptsize 0.43 & \scriptsize 0.12\\
\scriptsize 1.1 & \scriptsize 12.05 & \scriptsize 0.18 & \scriptsize 0.0175 & \scriptsize 0.0060 & \scriptsize 1.66 & \scriptsize 0.26 & \scriptsize 0.31 & \scriptsize 0.52 & \scriptsize 0.40\\
\scriptsize 1.5 & \scriptsize 12.15 & \scriptsize 0.30 & \scriptsize 0.0110 & \scriptsize 0.0044 & \scriptsize 1.29 & \scriptsize 0.25 & \scriptsize 0.32 & \scriptsize 0.41 & \scriptsize 0.41\\
\scriptsize 1.8 & \scriptsize 12.28 & \scriptsize 0.27 & \scriptsize 0.0116 & \scriptsize 0.0051 & \scriptsize 1.53 & \scriptsize 0.33 & \scriptsize 0.41 & \scriptsize 0.41 & \scriptsize 0.41\\
\scriptsize 2.5 & \scriptsize 12.22 & \scriptsize 0.38 & \scriptsize 0.0130 & \scriptsize 0.0037 & \scriptsize 0.90 & \scriptsize 0.20 & \scriptsize 0.24 & \scriptsize 0.30 & \scriptsize 0.30\\
\scriptsize 3.5 & \scriptsize 12.21 & \scriptsize 0.19 & \scriptsize 0.0101 & \scriptsize 0.0020 & \scriptsize 0.82 & \scriptsize 0.72 & \scriptsize 1.16 & \scriptsize 0.46 & \scriptsize 0.21\\
\enddata
\tablecomments{For $M_1$, $(m/M)_0$ and $\gamma$ the errors are drawn
  from a Gaussian and thus are symmetric (indicated by the symbol
  $\pm$). For $\beta$ the errors are drawn from a lognormal
  distribution and thus there is a lower error (indicated by the
  symbol $-$) and an upper error (indicated by the symbol $+$). All
  quoted masses are in units of $\msun$}
\label{t:pararedshifttab}
\end{deluxetable}

Having selected the observational SMFs for a set of
different redshifts, we fit the four free parameters $M_1$, $(m/M)_0$,
$\beta$ and $\gamma$ to the observations. The errors on the parameters
are derived in a similar way as explained in section \ref{s:probdis},
but instead of using confidence intervals we have fitted a Gaussian to
the probability distributions of $M_1$, $(m/M)_0$ and $\gamma$ and a
lognormal to the probability distribution of $\beta$.

Figure~\ref{f:fig12} shows the observed and the model stellar mass
functions for different redshifts (indicated at the top of each
panel). The values of the resulting four parameters for the different
redshifts are shown in Table~\ref{t:pararedshifttab} and the redshift
evolution is plotted in Figure~\ref{f:fig13}. The characteristic mass
$M_1$ grows with increasing redshift, while the normalization of the
SHM ratio $(m/M)_0$ becomes smaller with increasing redshift. This
means that there is less stellar content in a halo of a given mass at
a higher redshift.

The high mass slope $\gamma$ can be constrained only weakly. This is
due to the limitation of the available galaxy surveys. As the area of
the survey is small, the volume in which galaxies are detected is
limited, and thus massive galaxies are very rare. This results in
large error bars for the SMF for massive galaxies
which propagate into the error bars of $\gamma$.
The situation improves slightly for higher redshifts as the volume of higher
redshift bins is larger and thus more massive galaxies can be observed.
The value of $\gamma$ decreases with increasing redshift. For higher
redshifts ($z>1$) the error bars on $\gamma$ become very large because
of the limited area covered by the available deep surveys (in this case, GOODS).
We leave it up to the reader to assess the reliability of our results
at $z>1$ based on our quoted error bars.

The low mass slope $\beta$ seems to increase with redshift until
$z\approx2$ and then drops to a low value. However, as the redshift
increases it becomes more and more difficult to observe low mass
galaxies which are very faint. Thus the high redshift values for
$\beta$ are not very well constrained and perhaps not to be fully
trusted. We therefore assume that $\beta$ grows with increasing
redshift.

As we explained in Section~\ref{beta}, $\beta$ is strongly related to
the parameter $\alpha$ of the Schechter function. A small value of
$\beta$ corresponds to a large absolute value of $\alpha$ while a
large value of $\beta$ results in a low absolute value of
$\alpha$. This would mean that for higher redshifts the stellar mass
function would become shallower, in contradiction with observations
(e.g. \citealt{fontana2006} show that the absolute value of $\alpha$
increases with redshift). However, one has to remember that the halo
mass function also changes with redshift and becomes steeper. Thus the
halo mass function steepens more than the SMF, so
$\beta$ has to increase in order to compensate.

With the derived parameter values it becomes possible to interpolate
and find the SHM ratio at any redshift. This is done
by choosing a redshift-parameterization for each of the parameters.

As $M_1$ and $(m/M)_0$ do not change much above a redshift of $z>1.5$
we choose power laws for the redshift dependence:
\begin{equation}
\log M_1(z) = \log M_1\vert_{z=0} \cdot(z+1)^{\mu} \;.
\end{equation}
and 
\begin{equation}
\left( \frac{m}{M}\right)_0(z) = \left( \frac{m}{M}\right)_{z=0} \cdot(z+1)^{\nu} \;.
\end{equation}
with the normalizations $M_0$ and $(m/M)_{z=0}$ and the slopes $\mu$ and $\nu$.

To parameterize $\gamma$ over redshift, a linear dependence would lead
to a negative $\gamma$ at a certain redshift. Though this is not
forbidden, it leads to a SHM ratio which would be
increasing monotonically with halo mass which is inconsistent with
feedback processes at the massive end. Hence we also choose a
power-law parameterization for $\gamma$:
\begin{equation}
\gamma(z) = \gamma_0 \cdot(z+1)^{\gamma_1} \;.
\end{equation}
with the normalization $\gamma_0$ and the slope $\gamma_1$.

From Figure \ref{f:fig13} we are not able to infer whether $\beta$
converges to a constant value. Thus we adopt a simple linear
parameterization:
\begin{equation}
\beta(z) = \beta_1 \cdot z + \beta_0 \;.
\end{equation}
Note that we have also tried other parameterizations (constant
$\beta$, decreasing $\beta$) but could not reproduce the observed
stellar mass functions. Using the linear parameterization for $\beta$
and the power laws for the other parameters we were able to compute
stellar mass functions that are in good agreement with the observed
ones.

\begin{deluxetable}{l l l l l l l l l}
\tablecaption{Parameters for redshift dependent stellar-to-halo mass relation}
\tablehead{
  \colhead{} &
  \colhead{$M_1\vert_{z=0}$} &
  \colhead{$\mu$} &
  \colhead{$(m/M)_{z=0}$} &
  \colhead{$\nu$} &
  \colhead{$\gamma_0$} &
  \colhead{$\gamma_1$} &
  \colhead{$\beta_0$} &
  \colhead{$\beta_1$}
}
\startdata
 & 11.88 & 0.019 & 0.0282 & -0.72 & 0.556 & -0.26 & 1.06 & 0.17\\
$\pm$ & ~~0.01 & 0.002 & 0.0003 & ~0.06 & 0.001 & ~0.05 & 0.06 & 0.12\\
\enddata
\tablecomments{All quoted masses are in units of $\msun$\\}
\label{t:redpars}
\end{deluxetable}

A fit to the derived values presented in Table~\ref{t:pararedshifttab}
yields the parameters given in Table~\ref{t:redpars}. As we do not
fully trust the derived values of $\beta$ for $z\gtrsim2$ we neglect
these two values and fit a line to the remaining values of $\beta$.

\subsection{The stellar-to-halo mass relation for different redshifts} \label{s:diffred}

\begin{figure}
\centering
\epsscale{1.15}
\plotone{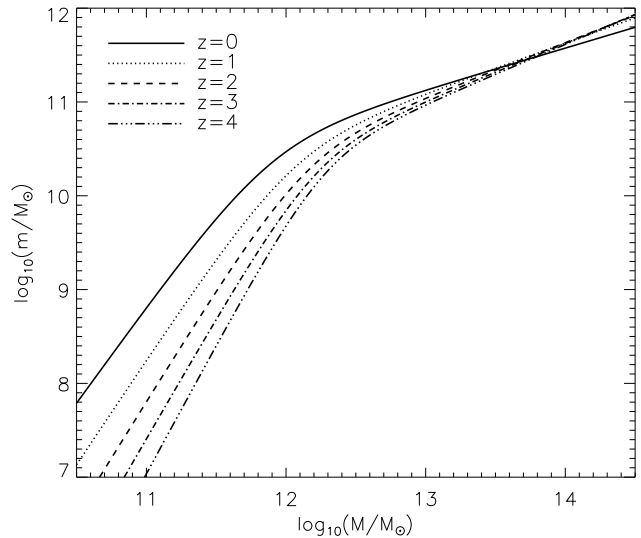}
\caption{Stellar mass as a function of halo mass for different
  redshifts. The solid lines show different redshifts, which are
  indicated at the top of the panels.}
\label{f:fig14}
\end{figure}

Having developed a redshift dependent model of the stellar-to-halo
mass relation we now test this model by computing interpolated stellar
mass functions for different redshifts. For this we use the method
described in section \ref{s:galaxies}. However, now we do not use the
parameters that have been derived at each redshift by fitting the
model to the observations but we use the eight parameters of the redshift
dependent SHM relation that have been derived in the previous section.

The resulting interpolated SMFs are compared to the
observations (and the fitted mass functions) in
Figure~\ref{f:fig12}. For $z\lesssim2$ we see excellent overall
agreement, the interpolated mass functions mostly overlap with the
error bars of the observations.

The SMFs for the high redshifts $z\gtrsim2$ are too low.
The deviations are largest at the low mass end. However, if we look at
Figure~\ref{f:fig12}, we see that $\beta$ is higher than the derived
value for the two highest redshifts which results in a low mass slope
that is too shallow.

\begin{figure*}
\centering
\epsscale{1.15}
\plotone{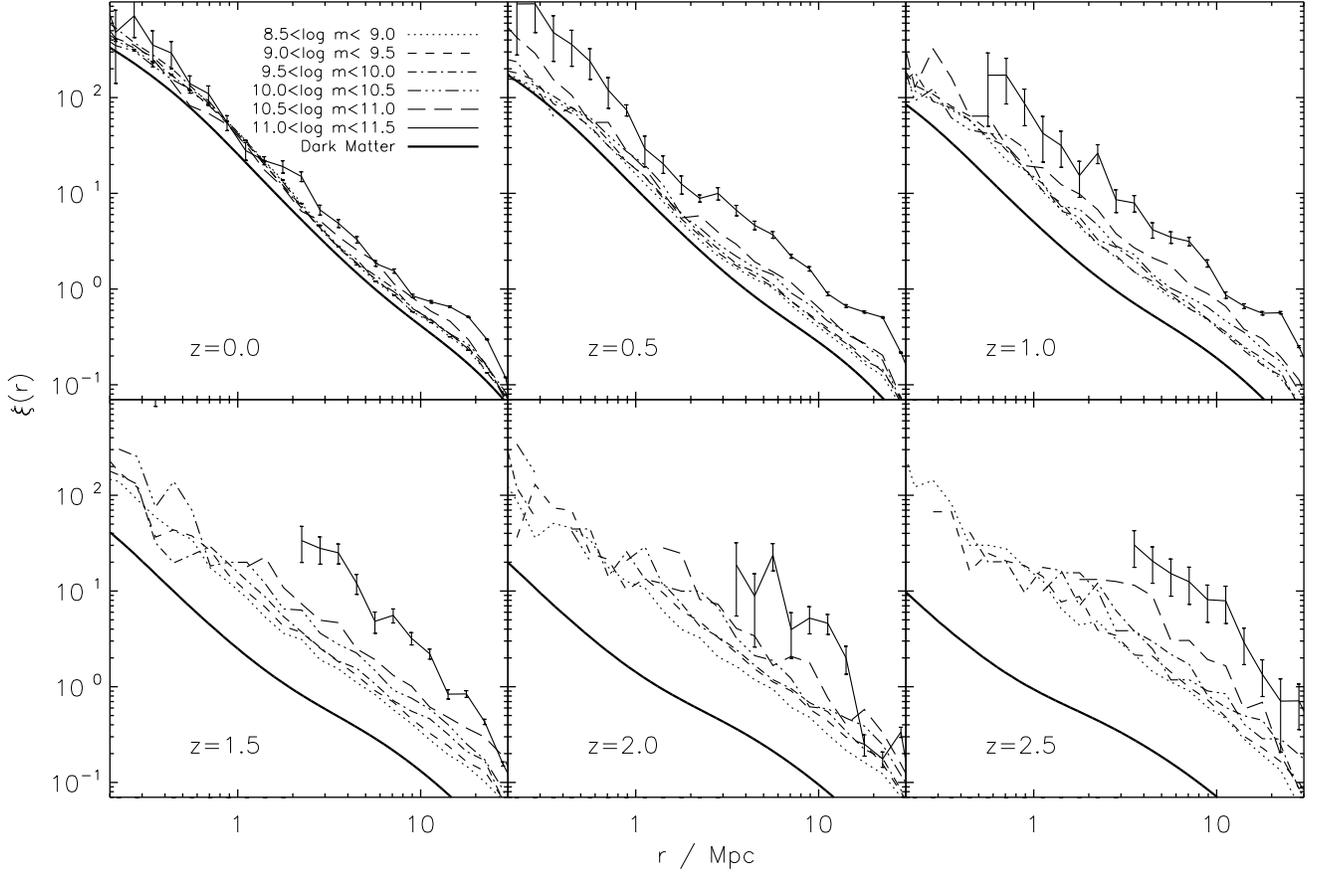}
\caption{Correlation functions as a function of stellar mass at high
  redshift. The different panels correspond to different redshifts,
  which are given at the bottom of each panel. The different lines are
  correlation functions for six stellar mass bins, which are given in
  the upper left panel. The error-bars on the most massive sample are
  from Poisson statistics. The correlation function of dark matter
  particles (thick solid line) at the respective redshifts is
  also shown for comparison.
  At high redshift the correlation function of the massive samples is only shown
  on large scales, since there is no relevant one-halo term.
  }
\label{f:fig15}
\end{figure*}

To compare the relation at different redshifts, we use the redshift
dependent SHM relation with the eight parameters that have been
derived in the previous section. Figure~\ref{f:fig14} plots stellar
mass versus halo mass for different redshifts. The plot shows that at
a fixed low halo mass (e.g. $M=10^{11}\msun$), galaxies that live in
such halos are more massive at low redshift ($m\sim10^9\msun$ for
$z=0$) than galaxies that live in a halo of the same mass at a higher
redshift ($m\sim10^8\msun$ for $z=2$). In contrast, massive halos
contain more massive galaxies at high redshift, while at low redshifts
the galaxies in massive halos have less mass. However, as halos also
become more massive over time, one cannot identify a halo of a certain
mass at high redshifts with a halo of the same mass at low
redshifts. Thus the fact that at a given (high) halo mass the mass of
the central galaxy is lower at present than at an earlier epoch does
not imply that individual galaxies lose mass during their
evolution. This only means that large halos accrete dark matter faster
than large galaxies grow in stellar mass, while the growth of low mass
halos is slower than that of the central galaxies they harbor
\citep[see also][]{conroy2008}. Because of its statistical nature, our
model is not suitable for following the evolution of an individual
galaxy through cosmic time. We also note that the SHM relation at the
massive end ($M\gta 10^{13}\msun$) undergoes very little evolution,
which has also been found by \citet{brown2008}.

\subsection{Clustering at higher redshift} \label{s:clustering}

\begin{figure}
\centering
\epsscale{1.15}
\plotone{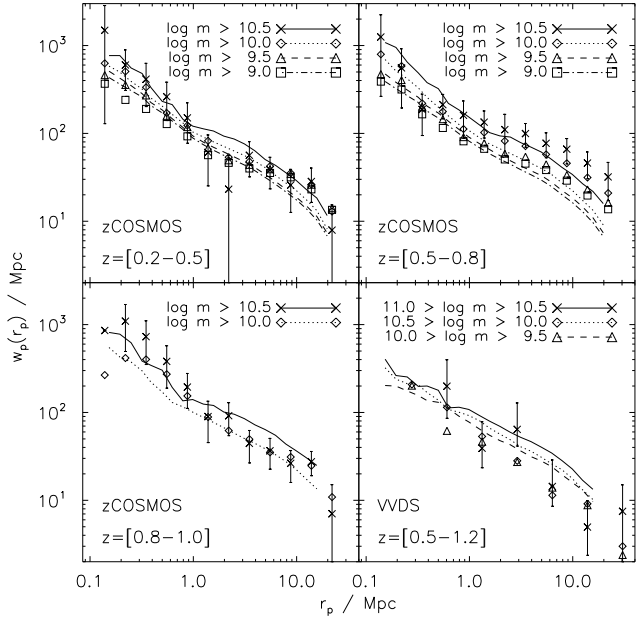}
\caption{Comparison between the model (lines) and observed (symbols)
   projected correlation functions at $0.2<z<1.2$. The upper and the left
   panels show the zCOSMOS data in three redshift bins while the lower
   right panel shows the VVDS data. The different lines and symbols in
   each panel are for different stellar mass bins and thresholds as indicated
   in the panels.}
\label{f:fig16}
\end{figure}

Having determined the SHM relation as a function of redshift we are
now able to populate halos with galaxies at any redshift. We choose a
set of redshifts and populate the halos with galaxies, deriving the
stellar masses from the redshift dependent SHM relation. We divide
these galaxies into six samples of different stellar mass between
$\log m/\msun=8.5$ and $11.5$. For each of these samples we compute
the real space CF $\xi(r)$ by counting pairs in distance bins
(equation \ref{corrfunc}). This leads to six CFs for every selected
redshift.

Figure~\ref{f:fig15} shows the CFs for six different redshifts as a
function of stellar mass. We also plot the correlation function of
dark matter at the respective redshifts for comparison. For all
redshifts we see that massive galaxies are clustered more strongly
than low mass galaxies. The higher the redshift, the more the CFs for
different stellar masses differ. For high redshift, there are very few
massive galaxies in our limited volume simulation box, and so the
error bars become larger.

At low redshift ($z\lta 1$), observational measurements of
stellar mass dependent galaxy clustering have recently been
published using the VIMOS-VLT Deep Survey (VVDS) and the zCOSMOS
Survey \citep{meneux2008,meneux2009}. In order to compare our model
predictions to these data, we compute correlation functions for the
same stellar mass bins and thresholds and convert these to projected
correlation functions as described in section
\ref{constraining}. Figure \ref{f:fig16} plots the observed
projected correlation functions (symbols) and the model predictions
(lines) for different stellar mass bins or thresholds in three
redshift bins for the zCOSMOS Survey and one redshift bin for the
VVDS. There is good general agreement between the model and
observations.  The zCOSMOS clustering amplitude agrees very well
with the model for $r_p<1{\rm~Mpc}$, but for $z<0.8$ deviates at
larger distances and becomes higher than the prediction. As
suggested by \citet{meneux2009}, this may be because the COSMOS
field represents an overdense volume at these redshifts. In
contrast, the VVDS clustering amplitudes are lower than those
predicted by our model, leading to the speculation that perhaps the
VVDS represents an underdense region.

\subsection{The galaxy bias} \label{bias}
\begin{figure}
\centering
\epsscale{1.05}
\plotone{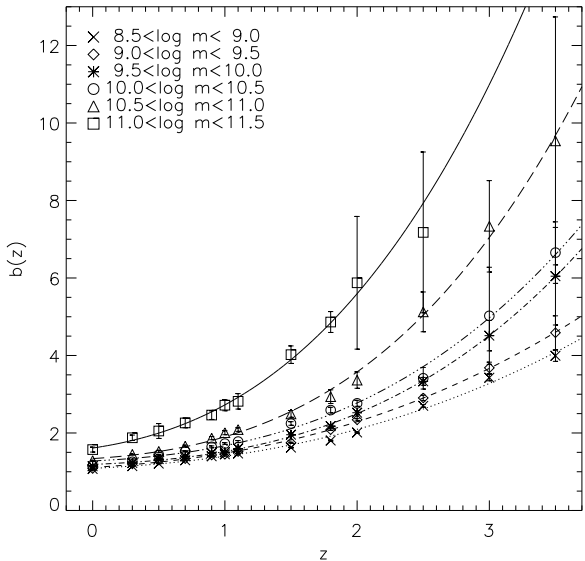}
\caption{The galaxy bias at a fixed scale ($\approx6{\rm~Mpc}$) as a
  function of redshift for different stellar masses. The symbols have
  been derived by averaging the bias over a distance interval while
  the lines are fits to the symbols.}
\label{f:fig17}
\end{figure}

The bias of any object may be defined as the square root of the ratio
between the CF of the object $\xi_o(r)$ and the CF of dark matter
particles $\xi_{dm}(r)$:
\begin{equation}
\label{bias1}
b(r)=\sqrt{\frac{ \xi_o(r) }{ \xi_{dm}(r) }}\quad.
\end{equation}
Here we focus on the galaxy two-point CF
$\xi_{gg}(r,m,z)$, which in addition to the distance between the
galaxies also depends on the redshift and the stellar mass of the
galaxies:
\begin{equation}
\label{bias2}
b(r,m,z)=\sqrt{\frac{ \xi_{gg}(r,m,z) }{ \xi_{dm}(r,z) }}\quad.
\end{equation}

From our predicted galaxy CFs, we compute the bias for every redshift
and stellar mass by averaging between $r=2{\rm~Mpc}$ and
$10{\rm~Mpc}$, where $b(r)$ is roughly a constant (as one can see from
Figure~\ref{f:fig15}, the scale dependence of the bias is quite
weak). Figure~\ref{f:fig17} shows the redshift dependence of the bias.
The symbols represent the averaged value of the bias while the
solid lines correspond to a fit to the symbols. For this we have used
a power law form:
\begin{equation}
b(z)=b_0 (z+1)^{b_1}+b_2 \quad
\end{equation}
where the parameters $b_0$, $b_1$, and $b_2$ are functions of stellar
mass. The fit parameters are given in Table~\ref{t:biasfit}.

\begin{deluxetable}{lccc}
\tablecaption{Galaxy bias fit parameters}
\tablehead{
  \colhead{$\log m_g$} &
  \colhead{$b_0$} &
  \colhead{$b_1$} &
  \colhead{$b_2$}
}
\startdata
~~8.5~-~~~9.0 & 0.062 $\pm$ 0.017 & 2.59 $\pm$ 0.18 & 1.025 $\pm$ 0.062\\
~~9.0~-~~~9.5 & 0.074 $\pm$ 0.008 & 2.58 $\pm$ 0.26 & 1.039 $\pm$ 0.028\\
~~9.5~-~10.0 & 0.042 $\pm$ 0.003 & 3.17 $\pm$ 0.05 & 1.147 $\pm$ 0.021\\
10.0~-~10.5 & 0.053 $\pm$ 0.014 & 3.07 $\pm$ 0.17 & 1.225 $\pm$ 0.077\\
10.5~-~11.0 & 0.069 $\pm$ 0.014 & 3.19 $\pm$ 0.13 & 1.269 $\pm$ 0.087\\
11.0~-~11.5 & 0.173 $\pm$ 0.035 & 2.89 $\pm$ 0.20 & 1.438 $\pm$ 0.061\\
\enddata
\tablecomments{All quoted masses are in units of $\msun$}
\label{t:biasfit}
\end{deluxetable}

This shows that the bias at a fixed stellar mass increases with
increasing redshift. Massive galaxies are biased more strongly
than galaxies of lower mass at any redshift. We find that the bias
of massive galaxies evolves more rapidly than that of low mass ones
\citep[cf.][]{white2007,brown2008}. Since the bias of massive halos evolves
more rapidly than that of low mass galaxies, this seems to be a
feature of any model in which the SHM relation is monotonically
increasing (i.e. the most massive galaxies reside in the most
massive halos).

\section{Conclusions}\label{s:conclusions}

The goal of this paper is to characterize the relationship between the
stellar masses of galaxies and the masses of the dark matter halos in
which they live at low and high redshift, and to make predictions of
stellar mass dependent galaxy clustering at high redshift.

We used a high-resolution $N$-body simulation and identified halos and
subhalos. Halos and subhalos were populated with central and satellite
galaxies using a parameterized SHM relation. For host halos the mass
was given by the virial mass $M_{\rm vir}$ while for subhalos we used
the maximum mass of the halo over its history $M_{\rm max}$ since we
expect the stellar mass of the satellite galaxy to be more tightly
linked to this quantity. 

We described the ratio between stellar and halo mass by a function
with four free parameters, a low-mass slope $\beta$, a characteristic
mass $M_1$, a high-mass slope $\gamma$, and a normalization
$(m/M)_0$. We fit for the values of these parameters by requiring that
the observed galaxy SMF is reproduced. We find that
the SHM function has a characteristic peak at
$M_1\sim 10^{12}\msun$, and declines steeply towards both smaller mass
($\beta \sim 1$) and less steeply towards larger mass halos ($\gamma \sim
0.6$). The physical interpretation of this behavior is the interplay
between the various feedback processes that impact the star formation
efficiency. Supernova feedback is more effective at reheating and
expelling gas in low mass halos, while AGN feedback is more effective
in high mass halos \citep[e.g.][]{shankar2006,croton06,bower2006,
somerville2008}. In this picture, the characteristic mass $M_1$ is the halo
mass where the efficiency of these two processes crosses.

We have thoroughly discussed the meaning of the parameters. We have
also investigated the effects on the SHM relation that arise from
introducing scatter to the relation. To do this we have added scatter
drawn from a lognormal distribution with a typical variance of
$\sigma_m=0.15$ to the SHM function. We showed that the impact of such
a scatter on three of the four parameters is negligible, with a small
but significant impact on the high-mass slope $\gamma$.

We showed that adding constraints from stellar mass dependent galaxy
clustering did not change the values of our best-fit parameters. Put
another way, the likelihood (here $\chi^2$) function for the
clustering constraint is much ``flatter'' than that for the mass
constraint, so adding the clustering constraint does not significantly
change the distribution for the most likely (best-fit) parameter
values.  Fitting to the SMF only, we found that the
observed projected CFs of galaxies for five samples
of different stellar mass were reproduced well. This means that the
clustering properties of galaxies are predominantly driven by the
clustering of the halos and subhalos in which they reside. From this
we concluded that our model can predict clustering as a function of
stellar mass at any redshift.
 
In order to describe how galaxies of different masses populate host
halos, we introduced the conditional mass function $\Phi(m\vert M)$,
which yields the average number of galaxies with stellar masses in the
range $m\pm{\rm d}m/2$ that live in a distinct halo of mass $M$. It is
described by five parameters which are functions of halo mass. We
divided the conditional mass function into a contribution from central
galaxies (described by a lognormal distribution) and a contribution
from satellite galaxies (described by a modified Schechter
function). We computed the SMF in different halo
mass bins and fitted the five parameters in each bin. Introducing halo
mass dependent functions for every parameter and fitting these to the
derived values of the parameters in the halo mass bins, we determined
the halo mass dependence of the five parameters and thus fully
described the conditional mass function. We also computed the
occupation numbers of halos which give the average number of galaxies
of a given stellar mass that live inside a halo of mass $M$.

We compared the results for our SHM function with
those that have been derived using other approaches. These include
other halo occupation type models, gravitational lensing and
semi-analytic models. We showed that all methods yield consistent
SHM relations.
 
Using SMFs at higher redshifts, we applied our model
at earlier epochs of the universe. We thus constrained the
SHM relation at a given set of redshifts between
$z=0$ and $z\sim4$. This allowed us to study how the four parameters
of the SHM function depend on redshift. For each
parameter we introduced a redshift dependent function. We found that
the characteristic mass increases with redshift while the
normalization decreases with redshift. This indicates that there is
less stellar content in halos at higher redshifts. 
As the halo mass function steepens more with redshift than the stellar
mass function, the low mass slope increases with redshift. We present
an eight parameter fitting function describing the redshift dependent
SHM relation.

Using the SHM relation that we derived in this way,
along with spatial information for halos from the $N$-body simulation,
we predicted the high-redshift real space CFs for
five stellar mass intervals.  We find that for all redshifts, massive
galaxies are more clustered than galaxies of lower mass. Using the
real space CF of dark matter we calculated the
galaxy bias as a function of distance, redshift and stellar
mass. Averaging over spatial scale in an interval around
$r\approx6\,\,{\rm Mpc}$, we demonstrated that the galaxy bias increases
with redshift, and presented fitting formulae for the galaxy bias as a
function of stellar mass and redshift. In a companion paper
\citep{mostercv} we will use these bias results to present predictions
for the cosmic variance $\sigma_c$ for galaxies of different stellar
mass.

\acknowledgments{
We thank Benjamin Panter, Cheng Li, Adriano Fontana, Niv Drory,
Rachel Mandelbaum and Baptiste Meneux for providing their data in
electronic form. We also thank Eric Bell, Ramin Skibba, Risa Wechsler
and Charlie Conroy for helpful discussions and Zheng Zheng,
Baptiste Meneux, Micheal Brown and the referee for helpful comments
on the draft version of this paper.
This research was supported in part by the DFG cluster
of excellence `Origin and Structure of the Universe'.
}

\clearpage

\end{document}